\documentclass[11pt, a4paper]{article}
\usepackage{styleBuding}
\usepackage{bm}
\usepackage{color}
\usepackage[usenames,dvipsnames,svgnames,table]{xcolor}
\usepackage{amsmath,amssymb}
\usepackage{graphicx}
\usepackage{slashed}
\usepackage{soul}
\usepackage{multirow}
\usepackage{subfigure}
\usepackage{indentfirst}
\usepackage{diagbox}
\usepackage{siunitx} % for physics units
\usepackage{url} %for link in @article
\usepackage[utf8]{inputenc}
\usepackage{cancel}

\pdfoutput=1
\usepackage{float}
\allowdisplaybreaks[4]

\begin{document}

\title{\boldmath Scalar Resonances near 650 and 95 GeV in the GNMSSM with Correct Dark Matter Relic Abundance}

%\author[1,2*]{Jingwei Lian\note[*]{Corresponding author.}}
%\author[1]{Yao-Bei Liu}
%\author[2,3]{Jinmin Yang}

%\author{Jingwei Lian$^{*1,2}$\note[*]{Corresponding author.}, Yao-Bei Liu$^1$, Jinmin Yang$^{2,3}$}

\author{Jingwei Lian$^{*1,2}$\note[*]{Corresponding author.}, Yao-Bei Liu$^1$}

\affiliation[1]{Henan Institute of Science and Technology, Xinxiang 453003, P.R. China}
\affiliation[2]{School of Physics, Henan Normal University, Xinxiang 453007, P.R. China}

\emailAdd{lianjw@hist.edu.cn}
\emailAdd{liuyaobei@hist.edu.cn}

\date{}

\abstract{%
Recent CMS analyses report an excess in the diphoton-plus-$b \bar{b}$ channel,  indicative of a heavy resonance around 650 GeV decaying into a Standard Model (SM)-like Higgs boson and a lighter scalar near 95 GeV. The case for a 95 GeV state is further supported by diphoton excesses observed by both CMS and ATLAS, as well as a $b\bar{b}$ excess previously observed at the Large Electron-Position collider. This study presents a unified interpretation of these anomalies within the framework of the General Next-to-Minimal Supersymmetric Standard Model that naturally accommodates a light singlet-dominated $CP$-even scalar boson $h_s$ near 95 GeV and a heavier doublet-like scalar boson $A_H$ near 650 GeV.
Through a comprehensive scan of the parameter space, we demonstrate that the model can explain these excesses at $2\,\sigma$ level while satisfying constraints from the dark matter relic density, direct detection experiments, the properties of the 125 GeV Higgs boson, $B$-physics observables, and searches for electroweakinos at the Large Hadron Collider (LHC).  
The interpretation features a Bino-dominated lightest neutralino as the dark matter candidate, whose relic abundance is achieved primarily via $A_s$ funnel annihilation or coannihilation with $\tilde{S}$-like $\tilde{\chi}^0_2$s into $h_sA_H$ final states.  Our findings provide clear predictions for testing this scenario at the high-luminosity LHC and future colliders.
}

\maketitle

\section{Introduction}
The monumental discovery of the Higgs boson at the Large Hadron Collider (LHC) in 2012 confirmed the existence of a scalar field responsible for electroweak symmetry breaking (EWSB) and mass generation. Although the measured properties of the 125 GeV Higgs boson are broadly consistent with Standard Model (SM) predictions, persistent theoretical issues—most notably the hierarchy problem—continue to motivate the exploration of physics beyond the Standard Model (BSM). Supersymmetry (SUSY) remains a compelling BSM framework, as it provides a natural solution to the hierarchy problem and typically necessitates an extended Higgs sector that predicts the existence of additional scalar states. The search for these new states is paramount, as their observation would provide essential insights into the structure of EWSB and the realization of BSM models.

Intriguingly, several collider experiments have reported persisting hints of such additional scalar states.
The first suggestion emerged from LEP experiments, where combined searches for the SM Higgs boson revealed a mild excess in the process $e^+e^- \rightarrow Z^* \rightarrow Z +b\bar{b}$ for Higgs masses in the range 95-100 GeV. This corresponded to a local significance of $2.3\,\sigma$ and a signal strength ~\cite{LEPWorkingGroupforHiggsbosonsearches:2003ing,Azatov:2012bz,Cao:2016uwt}:
\begin{equation}
\mu_{b \bar b}^\text{exp} = 0.117 \pm 0.057.  \label{LEPrate}
\end{equation}
This initial hint gained traction with a subsequent observation by CMS at the LHC, which reported an excess in the diphoton ($\gamma\gamma$) channel near 97 GeV during Run 1, carrying a local significance of approximately $2\,\sigma$~\cite{CMS:2015ocq}.  
Most recently, the case for this light scalar has been significantly reinforced using the full Run 2 dataset. Both the CMS and ATLAS collaborations, employing advanced analysis techniques, have observed a consistent excess at an invariant mass of $m_{\gamma\gamma} = 95.4$ GeV~\cite{CMS:2023yay, Arcangeletti}. 
The combined signal strength is now interpreted as a $3.1\,\sigma$ local excess with a signal rate~\cite{Biekotter:2023oen}:
\begin{equation}
\mu^{\rm exp}_{\gamma\gamma} \equiv  \mu^{\rm ATLAS + CMS}_{\gamma \gamma} = \frac{\sigma(p p \to \phi \to \gamma\gamma)}{\sigma_{\rm SM} (g g \to H_{\rm SM} \to \gamma\gamma)} = 0.24^{+ 0.09}_{-0.08},  \label{diphoton-rate}
\end{equation}
where, $\phi$ denotes a hypothetically non-SM scalar with mass $m_\phi = 95.4~{\rm GeV}$ responsible for the diphoton excess, and $\sigma_{\rm SM}$ refers to the cross section expected for an SM-like Higgs boson $H_{\rm SM}$ of the same mass.  
Additional phenomenological support for an about 95 GeV state arises from a mild surplus in the di-$\tau$ channel observed by CMS~\cite{CMS:2022goy} and a potential excess in $WW$ final states~\cite{Coloretti:2023wng}. 
Collectively, these observations strengthen the case for the existence of a light scalar resonance around 95 GeV. 

Complementing these hints at low mass, a recent search by CMS has unveiled an inter-connected anomaly — a $3.8\,\sigma$ local excess in the search channel featuring a diphoton a diphoton and a $b\bar{b}$ pair. This signal is attributed to the decay of a heavy resonance $X$ near 650 GeV into a SM-like Higgs boson and a BSM scalar $\phi$ in the 90–100 GeV mass range~\cite{CMS:2023boe}. The observed production cross section times branching fraction is~\cite{Ellwanger:2023zjc}:
\begin{equation}
 \sigma_{\gamma\gamma b\bar{b}} = \sigma(gg \to X \to H_{\rm SM}(\gamma\gamma) + \phi(b\bar{b})) = 0.35^{+0.17}_{-0.13} \;\rm fb . \label{XSbbrrExp}
\end{equation}
It is critical to consider constraints from complementary final states. Specifically, a prior CMS search in the $\tau\bar{\tau}$ plus $b\bar{b}$ channel \cite{CMS:2021yci} imposes a stringent upper $95\%$ CL limit of approximately $3\rm \;fb$ on the cross section $\sigma(X \to H_{\rm SM}(\tau\bar{\tau}) + \phi(b\bar{b}))$. This translates into an upper 95\% CL limit of about $0.1\rm \;fb$ on $\sigma_{\gamma\gamma b\bar{b}}$~\cite{Ellwanger:2023zjc}, suggesting a tension between the observed excess and current limits from related channels.

These collective anomalies have sparked extensive investigations across various BSM scenarios~\cite{ Fan:2013gjf, Cao:2016uwt, Biekotter:2017xmf, Beskidt:2017dil,  Fox:2017uwr,Haisch:2017gql,
Heinemeyer:2018wzl, Heinemeyer:2018jcd, Wang:2018vxp, Domingo:2018uim,Vega:2018ddp,
Cao:2019ofo,Biekotter:2019gtq, Choi:2019yrv,
Kundu:2019nqo,Biekotter:2019mib,Biekotter:2019kde,Sachdeva:2019hvk,
Biekotter:2020cjs,Abdelalim:2020xfk,Hollik:2020plc, Aguilar-Saavedra:2020wrj,
Biekotter:2021qbc, Biekotter:2021ovi,Heinemeyer:2021msz,
Biekotter:2022jyr,Li:2022etb, Benbrik:2022dja, Benbrik:2022azi, 
Li:2023hsr,Biekotter:2023oen,Biekotter:2023jld, Aguilar-Saavedra:2023vpd, Banik:2023ecr,Dutta:2023cig, Borah:2023hqw,Arcadi:2023smv,Ahriche:2023wkj, Chen:2023bqr,Dev:2023kzu,Ellwanger:2023zjc,Cao:2023gkc,Ahriche:2023hho,Belyaev:2023xnv, Azevedo:2023zkg, Ashanujjaman:2023etj,Karan:2023kyj,
Wang:2024bkg, Ellwanger:2024txc, Lian:2024smg, Cao:2024axg, Liu:2024cbr,Ellwanger:2024vvs, YaserAyazi:2024hpj, Gao:2024ljl, Gao:2024qag, Mondal:2024obd,Khanna:2024bah,Baek:2024cco,Dong:2024ipo, Kalinowski:2024uxe,
Xu:2025vmy, Abbas:2025ser, Li:2025tkm, Du:2025eop, Benbrik:2025hol,Benbrik:2025wkz,Khanna:2025cwq, Chang:2025bjt, Hmissou:2025riw, Dong:2025exu, Chen:2025vtg, Coutinho:2024zyp}.  
In particular, several works~\cite{Ellwanger:2023zjc, Banik:2023ecr, Azevedo:2023zkg, Benbrik:2025hol,Benbrik:2025wkz,Khanna:2025cwq} have explored the possible connection between the 95 GeV and 650 GeV excesses. 
Among the proposed explanations, the Next-to-Minimal Supersymmetric Standard Model (NMSSM)~\cite{Fayet:1974pd, Fayet:1976et, Fayet:1976cr, Fayet:1977yc, Farrar:1978xj, Ellwanger:2009dp, Maniatis:2009re} stands out as a particularly attractive and economical framework for accommodating these scalar states and furnishing proper dark matter (DM) candidates. 
Previous NMSSM studies with a $\mathbb Z_{3}$ symmetry have investigated both mass excesses~\cite{Ellwanger:2023zjc, Hammad:2025wst}, though typically under the assumption that the observed DM relic density receives additional contributions beyond the lightest supersymmetric particle (LSP). 
The general form of the NMSSM, termed as ``GNMSSM”, includes $\mathbb Z_{3}$-violating terms that naturally address the tadpole and domain-wall problems~\cite{Ellwanger:2009dp}. 
The enlarged parameter space, featuring ten parameters in the Higgs sector, offers the necessary flexibility to account for both the $95 \text{ GeV}$ and $650 \text{ GeV}$ excesses, while also providing a rich phenomenology for the Singlino/Bino DM sector. 
The $\mathbb Z_3$-breaking terms also modify the Higgsino and Singlino masses, significantly influencing the DM sector. Our recent work~\cite{Cao:2023gkc, Cao:2024axg} has shown that the 95~GeV excesses can be consistently explained in the GNMSSM through a Singlino-dominated LSP that saturates the observed DM relic density without violating existing experimental constraints.

The goal of this paper is to determine to what extent the 95~GeV and 650~GeV excesses can be simultaneously explained within the GNMSSM while remaining compatible with current LHC data and the recent null results from the LZ dark-matter experiment. In this framework, the heavy CP-even Higgs boson with mass near 650~GeV decays into a SM-like Higgs boson and a light singlet-dominated CP-even Higgs boson with mass around 95~GeV. We demonstrate that the GNMSSM can simultaneously account for the LEP, CMS, and ATLAS hints of a 95~GeV scalar as well as the CMS diphoton plus $b\bar b$ excess near 650~GeV at the $2\,\sigma$ level. We further identify parameter regions in which a Bino-dominated neutralino constitutes a viable thermal DM candidate, potentially testable by the forthcoming sensitivity of direct-detection experiments.

The remainder of this paper is organized as follows. In Sec.~\ref{Section-Model}, we briefly recapitulate the basics of the GNMSSM framework and the predicted signal rates for the anomalies. In Sec.~\ref{Section-explanation}, we perform a sophisticated scan over the model parameter space and present detailed numerical results. Finally, our conclusions are summarized in Sec.~\ref{conclusion}.

\section{Theoretical preliminaries}  \label{Section-Model}
\subsection{The Superpotential of the GNMSSM}
The NMSSM is known as a straightforward and well-motivated extension of the Minimal Supersymmetric Standard Model (MSSM), achieved by introducing a gauge-singlet chiral superfield~\cite{Fayet:1974pd, Fayet:1976et, Fayet:1976cr, Fayet:1977yc,Farrar:1978xj}. 
Its appeal stems not only from providing a natural solution to the so-called $\mu$-problem inherent in the MSSM, but also from offering an expanded and often more viable DM sector~\cite{Ellwanger:2014hia,Cao:2021ljw,Cao:2019qng}. 
Crucially, the NMSSM contributes to a sizeable mass lift for the SM-like Higgs boson.
Especially in scenarios involving a light $CP$-even Higgs, this mass can be elevated by both an additional tree-level contribution and a large singlet-doublet mixing~\cite{Ellwanger:2011aa,Badziak:2013bda,Cao:2012fz}, mitigating the need for substantial radiative corrections from top/stop loops. The GNMSSM incorporates the most general renormalizable couplings in its superpotential~\cite{Ellwanger:2009dp}
\begin{eqnarray}
 W_{\rm GNMSSM} = W_{\rm Yukawa} + \lambda \hat{S}\hat{H_u} \cdot \hat{H_d} + \frac{\kappa}{3}\hat{S}^3 + \mu \hat{H_u} \cdot \hat{H_d} + \frac{1}{2} \mu^\prime \hat{S}^2 + \xi\hat{S},  \label{Superpotential}
  \end{eqnarray}
where $W_{\rm Yukawa}$ contains the quark and lepton Yukawa terms in the MSSM superpotential, $\hat{H}_u=(\hat{H}_u^+,\hat{H}_u^0)^T$ and $\hat{H}_d=(\hat{H}_d^0,\hat{H}_d^-)^T$ are the $SU(2)_L$ Higgs doublet superfields, and $\hat{S}$ is the singlet superfield. 
The dimensionless parameters $\lambda$ and $\kappa$ play roles analogous to those in the $\mathbb{Z}_3$-invariant NMSSM, while the bilinear mass terms $\mu$ and $\mu^\prime$ and the singlet tadpole term $\xi$ explicitly break the $\mathbb{Z}_3$ symmetry. 
These $\mathbb{Z}_3$-breaking terms are advantageous, as they are essential for resolving the tadpole problem \cite{Ellwanger:1983mg, Ellwanger:2009dp} and the cosmological domain-wall problem that plagues the $\mathbb{Z}_3$-NMSSM \cite{Abel:1996cr, Kolda:1998rm, Panagiotakopoulos:1998yw}. 
Since one among $\mu$, $\mu'$, and $\xi$ may be eliminated through a shift of the $\hat{S}$ field followed by a redefinition of the remaining parameters~\cite{Ross:2011xv}, we set $\xi = 0$ without loss of generality. 
The bilinear parameters $\mu$ and $\mu^\prime$ may naturally reside at the electroweak scale, originating from an underlying discrete R-symmetry $Z^R_4$ after supersymmetry breaking \cite{Ross:2011xv,Lee:2010gv,Lee:2011dya,Ross:2012nr}. 

\subsection{Higgs Sector}
The soft SUSY-breaking Lagrangian for the Higgs fields in the GNMSSM takes the form:
\begin{equation}
\begin{aligned}
 -\mathcal{L}_{soft} = &\Bigg[\lambda A_{\lambda} S H_u \cdot H_d + \frac{1}{3} \kappa A_{\kappa} S^3+ m_3^2 H_u\cdot H_d + \frac{1}{2} {m_S^{\prime}}^2 S^2 + \xi^\prime S + h.c.\Bigg]   \\
& + m^2_{H_u}|H_u|^2 + m^2_{H_d}|H_d|^2 + m^2_{S}|S|^2 , \label{softL}
\end{aligned}
\end{equation}
Here, $H_u$, $H_d$, and $S$ denote the scalar components of the Higgs superfields, and $m^2_{H_u}$, $m^2_{H_d}$, and $m^2_{S}$ are their soft-breaking masses. 
After solving the EWSB conditional equations to minimize the scalar potential and expressing these soft masses in terms of the vacuum expectation values (vevs), namely, $\left\langle H_u^0 \right\rangle = v_u/\sqrt{2}$, $\left\langle H_d^0 \right\rangle = v_d/\sqrt{2}$, and $\left\langle S \right\rangle = v_s/\sqrt{2}$, with $v = \sqrt{v_u^2+v_d^2}\simeq 246~\mathrm{GeV}$ and $\tan{\beta} \equiv v_u/v_d$, the Higgs sector is governed by eleven independent parameters: $\tan{\beta}$, $v_s$, the Yukawa couplings $\lambda$ and $\kappa$, the soft-breaking trilinear coefficients $A_\lambda$ and $A_\kappa$,  the bilinear mass parameters $\mu$ and $\mu^\prime$, the soft-breaking parameters $m_3^2$ and $m_S^{\prime\ 2}$,  and the soft-breaking tadpole coefficient $\xi^\prime$ that is assumed to vanish the present analysis. 

To clearly delineate the properties of the Higgs states, it is convenient to use rotated field combinations: $H_{\rm SM} \equiv   \sqrt{2} {\rm Re} (\sin\beta  H_u^0 + \cos\beta H_d^0)$, $H_{\rm NSM} \equiv \sqrt{2} {\rm Re} ( \cos\beta H_u^0 - \sin\beta H_d^0)$, and $A_{\rm NSM} \equiv \sqrt{2} {\rm Im} ( \cos\beta H_u^0 + \sin\beta H_d^0)$, where $H_{\rm SM}$ behaves like the SM Higgs field, and $H_{\rm NSM}$ and $A_{\rm NSM}$ describing the additional doublet fields~\cite{Cao:2012fz}.  The singlet field remains unrotated and is written as $\sqrt{2} S \equiv H_{\rm S} + i A_S $.

In the basis $\left(H_{\rm NSM}, H_{\rm SM}, {\rm Re}[S]\right)$, the elements of the symmetric squared mass matrix for the $CP$-even Higgs boson  can be written in the following forms~\cite{Ellwanger:2009dp, Miller:2003ay}:
\begin{equation}
\begin{aligned}
  {\cal M}^2_{S, 11}&= m^2_A + \frac{1}{2} (2 m_Z^2- \lambda^2v^2)\sin^22\beta,  \\
  {\cal M}^2_{S, 12}&=-\frac{1}{4}(2 m_Z^2-\lambda^2v^2)\sin4\beta, \\
  {\cal M}^2_{S, 13}&= \sqrt{2} \lambda \mu_{tot} v (\delta -1) \cot{2\beta},  \\
  {\cal M}^2_{S, 22}&=m_Z^2\cos^22\beta+ \frac{1}{2} \lambda^2v^2\sin^22\beta,  \\
  {\cal M}^2_{S, 23}&= \sqrt{2} \lambda \mu_{tot} v \delta, \\
  {\cal M}^2_{S, 33}&= m_B^2,
 \label{CPevenHiggsMass}
\end{aligned} 
\end{equation}
while those for the $CP$-odd Higgs boson in the basis $\left(A_{\rm NSM}, A_S\right)$ are:
\begin{eqnarray}
{\cal M}^2_{P,11} = m_A^2, 
 \quad {\cal M}^2_{P,22} &=& m_C^2, \quad {\cal M}^2_{P,12} = \frac{\lambda v}{\sqrt{2}} ( A_\lambda - m_N ).
 \label{CPoddHiggsMass}
 \end{eqnarray}
We {  {introduce}} a $\delta$ factor in ${\cal M}^2_{S, 13}$ and ${\cal M}^2_{S, 23}$, defined as:
\begin{equation}
 \delta \equiv 1- \frac{(A_\lambda + m_N)\sin2\beta}{2\mu_{tot}}, \label{delta}
 \end{equation}
with the Higgsino mass $\mu_{tot} \equiv \mu_{eff}+\mu = \lambda v_s/\sqrt{2} + \mu$ and the Singlino mass $m_N \equiv  \sqrt{2} \kappa v_s + \mu^\prime$, to simplify the expression and allow for direct manipulation of the mixings of $H_{\rm S}$ with $H_{\rm SM}$ and $H_{\rm NSM}$. This factor is expected to be tiny in the alignment without decoupling limit~\cite{Carena:2015moc, Biekotter:2021qbc}. 
$m_A$ is the typical definition of the mass of the doublet-like pseudoscalar Higgs field. $m_B$ and $m_C$ characterize the singlet-like CP-even, and the singlet-like CP-odd Higgs masses, respectively.  Then the soft-breaking parameters $A_\kappa$, $m_3^2$ and $m_S^{\prime\ 2}$ can be expressed in terms of the physical input parameters:
\begin{eqnarray} 
A_\kappa &=& \frac{2 m^2_B}{m_N-\mu^\prime} + \frac{\lambda^2v^2}{4\kappa\mu^2_{eff}} \left[ (A_\lambda + \mu^\prime)\sin 2\beta - 2\mu \right] - 2m_N - \mu^\prime, \nonumber\\
m^2_3 &=& \frac{1}{2}\left[ m^2_A \sin{2\beta} - \mu_{eff}(2A_\lambda + m_N + \mu^\prime) \right], \nonumber \\
m_S^{\prime 2} &=& \frac{1}{2} \left[ m_B^2 - m_C^2 + \lambda \kappa v^2 \sin 2\beta - (m_N - \mu^\prime) ( 2A_\kappa + m_N + \mu^\prime) \right].
\label{Simplify-1}
\end{eqnarray} 
Note that the mass parameter for the singlet-like $CP$-odd Higgs $m_C$ is fixed at $800 \rm GeV$ in this study for simplicity to be consistent with current collider experimental limitations ~\cite{ATLAS:2020zms,CMS:2022goy}.  

Diagonalizing the squared mass matrix with mixing angles, denoted as $V^j_{h_i}$, yields three $CP$-even Higgs mass eigenstates:
  \begin{eqnarray}
    h_i & = & V_{h_i}^{\rm NSM} H_{\rm NSM}+V_{h_i}^{\rm SM} H_{\rm SM}+V_{h_i}^{\rm S} H_{\rm S},
   \label{Vij}
  \end{eqnarray}   
with $h_i = \{ h_{\rm s}, h, H \}$ ordered by ascending mass. The second $CP$-even mass eigenstate $h$ corresponds to the observed 125 GeV SM-like Higgs boson. This state is predominantly composed of the $H_{\rm SM}$ component, requiring contributions from $H_{\rm NSM}$ and Re$[S]$ components restricted to be less than about 10\% \cite{ATLAS:2022vkf,CMS:2022dwd}, i.e., $\sqrt{\left (V_h^{\rm NSM} \right )^2 + \left ( V_h^{\rm S} \right )^2} \lesssim 0.1$ and $|V_h^{\rm SM}| \sim 1$ consistent with current Higgs data. In addition, the imaginary components $A_S$ and $A_{\rm NSM}$ mix into two $CP$-odd Higgs mass eigenstates $a_i=\{A_H, A_s\}$, while the charged components give rise to a pair of charged Higgs bosons $H^\pm$. The light $CP$-odd Higgs boson $A_H$ and the charged Higgs bosons $H^\pm$ should exhibit a nearly degenerate mass with the heavy $CP$-even boson $H$.

The present study assumes that the lightest $CP$-even Higgs boson $h_s$ is singlet-dominated and fully accounts for both the $\gamma\gamma$ and $b\bar{b}$ excesses near 95.4 GeV. The signal strengths normalized to their SM predictions  are~\cite{Cao:2023gkc}:
\begin{eqnarray}
	\mu_{\gamma\gamma}|_{m_{h_s} = 95.4~{\rm GeV}} &=&
  \frac{\sigma_{\rm SUSY}(p p \to h_s)}
       {\sigma_{\rm SM}(p p \to h_s )} \times
       \frac{{\rm Br}_{\rm SUSY}(h_s \to \gamma \gamma)}
       {{\rm Br}_{\rm SM}(h_s \to \gamma \gamma)},  \label{muCMS}  \\
  	\mu_{b\bar{b}}|_{m_{h_s} = 95.4~{\rm GeV}} &=&
  \frac{\sigma_{\rm SUSY}(e^+e^-\to Z h_s)}
       {\sigma_{\rm SM}(e^+e^-\to Z h_s)} \times
       \frac{{\rm Br}_{\rm SUSY}(h_s\to b\bar{b})}
       {{\rm Br}_{\rm SM}(h_s \to b\bar{b})},  \label{muLEP}
  \end{eqnarray}
where the mass of $h_s$ is fixed at $95.4~{\rm GeV}$. The production rate $\sigma( p p \to h_s)$ and the decay branching ratio ${\rm Br} (h_s \to \gamma \gamma)$ labeled with the subscript `SUSY' refer to the predictions from the model, whereas those with the subscript `SM' assume SM couplings for $h_s$. 

The $\gamma\gamma b\bar{b}$ excess near 650 GeV can be attributed to the heavy doublet-dominated $CP$-even Higgs boson $H$ decaying into a SM-like Higgs boson $h$ and a singlet-dominated Higgs boson $h_s$. The cross section of the resonance in the $\gamma\gamma b\bar{b}$ channel can be expressed as follows:
\begin{equation}
 \sigma_{\gamma\gamma b\bar{b}} = \sigma(gg \to H) \times {\rm Br}_{\rm SUSY}(H \to h h_s) \times {\rm Br}_{\rm SUSY}(h \to \gamma \gamma) \times {\rm Br}_{\rm SUSY}(h_s \to b \bar{b}),  \label{XSbbrr}
\end{equation}
where the masses of $H, h$ ad $h_s$ are fixed at approximately  650 GeV, 125 GeV and 95.4 GeV, respectively.  
By fitting the GNMSSM predictions for $\mu_{\gamma\gamma}$, $\mu_{b\bar{b}}$, and $\sigma_{\gamma\gamma b\bar{b}}$ to the observed experimental rates, we can investigate the viability of the GNMSSM as a unified interpretation for the $95 \text{ GeV}$ and $650 \text{ GeV}$ excesses.

\subsection{Neutralino Sector}
The neutralino sector within the GNMSSM framework is comprised of five superpartners: 
the gaugino fields ($\tilde{B}$, the Bino, and $\tilde{W}$, the Wino), the Higgsino fields ($\tilde{H}_d^0$ and $\tilde{H}_u^0$), and the Singlino field ($\tilde{S}$).
In the basis $\psi^T \equiv (\tilde{B},\tilde{W},\tilde{H}_d^0,\tilde{H}_u^0,\tilde{S})$, the symmetric $5 \times 5$ neutralino mass matrix, $\mathcal{M}$, is given by~\cite{Ellwanger:2009dp}:
  \begin{equation}
    {\cal M} = \left(
    \begin{array}{ccccc}
    M_1 & 0 & -m_Z \sin \theta_W \cos \beta & m_Z \sin \theta_W \sin \beta & 0 \\
      & M_2 & m_Z \cos \theta_W \cos \beta & - m_Z \cos \theta_W \sin \beta &0 \\
    & & 0 & -\mu_{tot} & - \frac{1}{\sqrt{2}} \lambda v \sin \beta \\
    & & & 0 & -\frac{1}{\sqrt{2}} \lambda v \cos \beta \\
    & & & & m_N
    \end{array}
    \right), \label{eq:MN}
    \end{equation}
where $\theta_W$ is the weak mixing angle, and $M_1$ and $M_2$ are the soft-breaking masses of the Bino and Wino, respectively. Diagonalizing $\cal{M}$ by a rotation matrix $N$ yields five neutralino mass eigenstates:
\begin{equation}
\tilde{\chi}_i^0 = \sum_{j=1}^5 N_{ij}\,\psi^0_j,
\qquad i=1,\ldots,5,
\end{equation}
ordered by increasing mass.  
The element $N_{ij}$ parametrizes the contribution of the interaction eigenstate $\psi^0_j$ to the physical neutralino $\tilde{\chi}_i^0$.
Under $R$\,parity conservation, the lightest neutralino $\tilde{\chi}_1^0$ is stable and constitutes a viable DM candidate when it is the Lightest Supersymmetric Particle (LSP).  
In many NMSSM realizations, a Singlino-dominated LSP naturally reproduces the observed relic abundance while evading stringent direct detection bounds~\cite{Baum:2017enm}.  
In the present study, however, this possibility is strongly limited.  
The relatively large value of $\lambda$ required to account for both the excesses enhances the Higgs-mediated scattering cross section of a Singlino-like LSP, bringing it into tension with current direct detection limits~\cite{Cao:2019qng, Meng:2024lmi}.  
Furthermore, the mixing constraints on $V_{h_i}^{\rm SM}$ and $V_{h_i}^{\rm NSM}$ required by the 95 GeV and 650 GeV anomalies further reduce the viable parameter space for Singlino-like DM.
In contrast, as will be demonstrated later, the combined interpretation of the excesses instead favors a Bino-dominated LSP.  
Such a Bino-like neutralino can efficiently reduce its relic abundance through Higgs-funnel annihilation and/or coannihilations with Singlino-, Higgsino-, or Wino-like electroweakinos.
The spin-independent (SI) and spin-dependent (SD) $\tilde{\chi}_1^0$-nucleon scattering cross-sections depend on the $\tilde{\chi}_1^0$ couplings to the Higgs boson and the $Z$ gauge boson, respectively. Both can be heavily suppressed due to cancellations among different contributions or by the emergence of blind-spot conditions~\cite{Huang:2014xua, Crivellin:2015bva, Han:2016qtc, Carena:2018nlf, Abdallah:2020yag,Chatterjee:2022pxf,Datta:2022bvg,Roy:2024yoh,Arganda:2025fhx,Zhou:2025xol}.

\section{Explanation of the excesses}  \label{Section-explanation}
This section outlines the parameter-sampling strategy and presents detailed numerical results that account for the observed excesses in the diphoton and $b\bar{b}$ final states—both separately and simultaneously—while remaining consistent with other experimental constraints, in particular those from Higgs data and DM detections.  
The numerical procedure follows several steps.  
First, we implement the GNMSSM using \textsf{SARAH-4.14.3}~\cite{SARAH_Staub2008,SARAH3_Staub2012,SARAH4_Staub2013,SARAH_Staub2015} to generate analytical model files. 
Second, the packages \textsf{SPheno-4.0.5}~\cite{Porod2003SPheno,Porod2011SPheno3} and \textsf{FlavorKit}~\cite{Porod:2014xia} are used to compute the SUSY mass spectrum and low-energy flavor observables, respectively.  
Subsequently, DM observables—including relic density, annihilation channels, and direct detection rates—are evaluated with \textsf{MicrOMEGAs-5.0.4}~\cite{Belanger2002,Belanger2004,Belanger2005,Belanger2006,BelangerRD2006qa,Belanger2008,Belanger2010pz,Belanger2013,Barducci2016pcb,Belanger2018}. 
Finally, the resulting samples are then analyzed using both Bayesian and Frequentist statistical frameworks: the posterior probability density function (PDF) and the profile likelihood (PL)~\cite{Fowlie:2016hew}.
These complementary approaches provide a robust characterization of the viable parameter space and enable a nuanced interpretation of the experimental anomalies.

\subsection{Strategy in scanning the parameter space} \label{scanStrategy}
To thoroughly investigate the phenomenological implications of the observed excesses and DM constraints, we employ a sophisticated scan strategy focusing on phenomenologically relevant input parameters.
The scan includes all parameters relevant to the Higgs sector. However, instead of directly scanning the original soft-breaking parameters ($m^2_3, A_\kappa, A_\lambda$) appearing in Eq.~(\ref{softL}), we trade them for the physical and semi-physical masses and mixing terms ($m_A$, $m_B$, and $\delta$) as defined in Eqs.~(\ref{Simplify-1}) and~(\ref{delta}).
This substitution improves the efficiency of the sampling by allowing direct exploration of regions consistent with a target Higgs mass spectrum while preserving the essential physics.
The soft trilinear couplings of the third-generation squarks, $A_t$ and $A_b$, are treated as free parameters because of their crucial impact on the radiative corrections to the SM-like Higgs mass.
Furthermore, the soft gaugino masses $M_1$ and $M_2$ are included as scan parameters to allow for a Bino-dominated $\tilde{\chi}_1^0$ DM candidate, as suggested by the analysis in Sec.~\ref{Section-Model}.
In total, the scan explores a 12-dimensional parameter space as delineated in Table.~\ref{tab:scan}. The parameter ranges are carefully determined based on theoretical considerations and exploratory trial scans across broader intervals to ensure both comprehensive coverage and effective exploration of the regions of interest. 
All final samples are required to respect the constraint of field-theoretic perturbativity between the electroweak and GUT scales. 
The approximate bound derived for the $\mathbb{Z}_3$-NMSSM~\cite{Miller:2003ay}, $\lambda^2 + \kappa^2 \lesssim 0.5$, is applied here, as the GNMSSM beta functions do not introduce new scale-dependent contributions up to the two-loop level (see Appendices in Refs.~\cite{Ellwanger:2009dp, King:1995vk, Masip:1998jc}). 
All numerical results presented in this work are ensured to satisfy this condition.

\begin{table}[tbp]
\caption{Ranges of input parameters used in this study. All parameters are assigned flat priors due to their well-defined physical interpretation. The soft trilinear coefficients for third-generation squarks are unified as $A_t = A_b$. Dimensional parameters not directly relevant here are fixed to simplify the analysis and comply with experimental constraints: the gluino mass is set to $M_3 = 3~\rm TeV$, and other unspecified parameters are fixed at $2~\rm TeV$ to satisfy LHC limits on SUSY particles. All parameters are defined at the renormalization scale $Q_{\rm input} = 1~\rm TeV$.
\label{tab:scan}}
\centering
\vspace{0.3cm}
\resizebox{0.7\textwidth}{!}{
\begin{tabular}{c|c|c|c|c|c}
\hline\hline
Parameter & Prior & Range & Parameter & Prior & Range   \\
\hline
$\lambda$ & Flat & $0.5 \sim 0.7$ & $\mu_{tot}/{\rm GeV}$ & Flat & $500.0 \sim 1000 $ \\
$\kappa$ & Flat & $-0.5 \sim 0.5$ & $\mu_{eff}/{\rm GeV}$ & Flat & $-1000 \sim 1000$ \\
$\delta$ & Flat & $-0.05 \sim 0.05$ & $m_A/{\rm GeV}$ & Flat & $500.0 \sim 650.0$\\
$\tan \beta$ & Flat & $1.0 \sim 2.0$ & $m_B/{\rm GeV}$ & Flat & $90.0 \sim 120$ \\
$M_1/{\rm GeV}$ & Flat & $-1000 \sim -200.0$ &  $A_t/{\rm GeV}$ & Flat & $1000 \sim 3000$ \\
$M_2/{\rm GeV}$ & Flat & $200.0 \sim 1000$ &  $m_N/{\rm TeV}$ & Flat & $-1000 \sim 1000$ \\
\hline\hline
\end{tabular}}
\end{table}

The multi-dimensional parameter space is explored using the MultiNest algorithm \cite{MultiNest2009,Importance2019}, a highly efficient nested sampling method, with a fixed number of live points set to ${\it{nlive}} = 6000$ to ensure comprehensive survey and accurate computation of the Bayesian evidence and posterior distributions\footnote{In nested sampling algorithms, \textit{nlive} denotes the number of active (live) points used to define successive iso-likelihood contours~\cite{MultiNest2009,Importance2019}. Larger values of \textit{nlive} improve the resolution of the parameter space at the cost of increased computational time.}. 
The likelihood function guiding the scan is constructed as
$\mathcal{L} \equiv  \mathcal{L}_{650 + 95} \times \mathcal{L}_{\rm Res}$. The first factor $\mathcal{L}_{650 + 95} = \exp ( -\frac{1}{2}\chi^2_{650 + 95} )$ quantifies the compatibility of the GNMSSM predictions with the central values and uncertainties of the experimental observations of the $\gamma\gamma$ plus $b\bar{b}$ excesses near $650{\rm GeV}$ and the $95{\rm GeV}$ diphoton and $b\bar b$ excesses, via a $\chi^2$ function:
\begin{equation}
\begin{split}
\chi^2_{650 + 95} = \left( \frac{ \sigma_{\gamma\gamma b\bar{b}} - 0.35~{  {\rm fb}}}{0.13~{  {\rm fb}} }\right)^2 + \left( \frac{\mu_{\gamma\gamma} - 0.24}{0.08}\right)^2 + \left( \frac{\mu_{b\bar{b}} - 0.117}{0.057}\right)^2 .   
\end{split}
\label{chi2-excesses}
\end{equation}
The second factor, $\mathcal{L}_{\rm Res}$ term enforces essential physical and experimental bounds by acting as a step function: $\mathcal{L}_{\rm Res} = 1$ if all conditions are met, and $\mathcal{L}_{\rm Res} = \exp\left [-100 \right ]$ otherwise. The required restrictions include:
\begin{itemize}
\item \textbf{Masses of the three scalars:} 
To interpret the observed excesses, we impose specific mass requirements on the three $CP$-even Higgs bosons:
\begin{itemize}
\item Light Scalar ($h_s$): {  {The mass of the light scalar boson is required to be in the range $95.4 \pm 1 \text{ GeV}$ to align with experimental hints. This choice is motivated by the fact that a singlet-like boson typically receives smaller radiative corrections than the SM-like Higgs.}}
\item Heavy Resonance ($H$): The heavy scalar mass is constrained to the range $650 \pm 25 \text{ GeV}$ to match the resonant excess observed by CMS~\cite{Ellwanger:2023zjc, CMS:2023boe}.
\item SM-like Higgs ($h$): The mass of the SM-like Higgs boson must be consistent with the LHC measurements, allowing for a $\pm 3 \text{ GeV}$ tolerance to account for combined experimental and theoretical uncertainties.
\end{itemize}
\item \textbf{SM-like Higgs data fit:} 
The SM-like Higgs boson must satisfy the ATLAS and CMS measurements at the $95\%$ confidence level. This condition is validated using the newest version of the \textsf{HiggsTools} code that incorporates the \textsf{HiggsSignals-2}~\cite{HS2013xfa,HSConstraining2013hwa,HS2014ewa,HS2020uwn}, which compares model predictions to a full set of current measurements of the 125GeV Higgs properties and returns a $\chi^2$ value that quantifies the agreement of the model predictions with the measurements. 
{  { The $\chi^2$ value is directly incorporated into the likelihood function. When presenting results in two-dimensional parameter planes, samples satisfying $\Delta\chi^2_{125} \equiv \chi^2 - \chi^2_{125,\mathrm{SM}} \lesssim 6.18$, where $\chi^2_{125,\mathrm{SM}} = 153$ denotes the SM best-fit value, are considered consistent with the Higgs data at approximately the $2\,\sigma$ confidence level (assuming Gaussian uncertainties)~\cite{Muhlleitner:2020wwk}. }}
\item \textbf{Extra Higgs searches:} 
Signal rates for all non-SM Higgs bosons ($h_s, H, A_H, H^\pm$) must comply with cross-section limits based on experimental data at LEP, Tevatron, and the LHC. 
This requirement is implemented using the \textsf{HiggsTools} code~\cite{Bahl:2022igd} with the complete \textsf{HiggsBounds-5.10.2} library~\cite{HB2008jh,HB2011sb,HBHS2012lvg,HB2013wla,HB2020pkv}.
\item \textbf{DM relic density:} The predicted DM relic density must agree with the Planck-2018 central value ($\Omega {h^2}=0.120$) \cite{Planck:2018vyg}, allowing for a conservative $20\%$ theoretical uncertainty: $0.096 \leq \Omega {h^2} \leq 0.144$.
\item \textbf{DM detections:} The SI and SD $\tilde{\chi}_1^0$-nucleon scattering cross-sections must be below the currently most stringent upper bounds imposed by the LZ experiments \cite{LZ:2024zvo, LZ2024slides}. Constraints from DM indirect searches are not considered here as they are less restrictive for the masses under consideration ($|m_{\tilde{\chi}_1^0}| \gtrsim 100 \text{ GeV}$) \cite{Fermi-LAT:2015att}.
\item \textbf{$B$-physics observables:} The branching ratios $B_s \to \mu^+ \mu^-$ and $B \to X_s \gamma$ are required to agree with current experimental measurements at the $2\,\sigma$ level~\cite{ParticleDataGroup:2024cfk}.
\item \textbf{Vacuum stability:} The electroweak vacuum must be demonstrated to be either the global minimum or a sufficiently long-lived, metastable vacuum state \cite{Hollik:2018wrr}. This is rigorously tested using the \textsf{VevaciousPlusPlus} code \cite{VPP2014,Camargo-Molina:2013qva}.
\end{itemize}

Constraints from LHC searches for electroweakinos are expected to have a negligible impact on the scan results.
This is because the gaugino–Higgsino states in Table.~\ref{tab:scan} lie in relatively heavy mass ranges, and all other SUSY particles are assumed to be heavy and decoupled. 
This estimation is consistent with recent statistical combinations of ATLAS Run 2 searches \cite{ATLAS:2024qxh}. 
To confirm this, we evaluate all sampled points using \texttt{SModelS-3.0.0}, which implements a wide set of simplified-model constraints by decomposing SUSY signatures into experimentally tested topologies and applying the corresponding efficiency maps~\cite{Khosa:2020zar}.
Furthermore, four representative benchmark points are subjected to dedicated Monte Carlo simulations.
For each benchmark, $10^6$ events were generated with \texttt{MadGraph\_aMC@NLO}~\cite{Alwall:2011uj, Conte:2012fm} for the following electroweakino production channels: 
\begin{equation}\begin{split}
pp \to \tilde{\chi}_i^0\tilde{\chi}_j^{\pm} &, \quad i = 2, 3, 4, 5; \quad j = 1, 2 \\
pp \to \tilde{\chi}_i^{\pm}\tilde{\chi}_j^{\mp} &, \quad i,j = 1, 2; \\
pp \to \tilde{\chi}_i^{0}\tilde{\chi}_j^{0} &, \quad i,j = 2, 3, 4, 5. 
\end{split}\end{equation} 
Production cross sections are computed at next-to-leading order using \texttt{Prospino2}~\cite{Beenakker:1996ed}.
Parton showering and hadronization are performed with \texttt{PYTHIA8}~\cite{Sjostrand:2014zea}, and detector effects are simulated using \texttt{Delphes}~\cite{deFavereau:2013fsa}.
The resulting events are analyzed with \texttt{CheckMATE-2.0.26}~\cite{Drees:2013wra,Dercks:2016npn,Kim:2015wza}, which computes the $R$ value, defined as the exclusion ratio $R \equiv max\{S_i/S_{i,obs}^{95}\}$, with $S_i$ representing the simulated event number of the $i$-th signal region (SR), and $S_{i,obs}^{95}$ the corresponding $95\%$ confidence level upper limit.
Values of $R > 1 $ indicates exclusion (neglecting theoretical and experimental uncertainties)~\cite{Cao:2021tuh}, while $R < 1$ denotes consistency with current experimental analyses.
The specific ATLAS and CMS analyses used in this study are listed in Table.~\ref{LHCanalyses} in the Appendix.

\subsection{Numerical Results}
The scan process yields over 25,000 parameter samples that satisfy all applied theoretical and experimental constraints. 
A key result across all viable samples is the preference for the Bino-dominated $\tilde{\chi}^0_1$ as the DM candidate. 
These samples naturally cluster into two distinct scenarios according to the dominant mechanism responsible for achieving the observed relic abundance.:   
 \begin{itemize}
    \item {\bf Scenario I} ($\approx 92\%$): Achieves the correct relic density primarily via $A_s$ funnel annihilation or coannihilation with $\tilde{S}$-like $\tilde{\chi}^0_2$s, typically into the $h_sA_H$ final state.
    \item {\bf Scenario II} ($\approx 8\%$):  Achieves the correct relic density through coannihilation with $\tilde{H}$-like $\tilde{\chi}^0_2$s or $\tilde{\chi}^\pm_1$s, favoring the $h_sA_s$ final state.
 \end{itemize}

\begin{figure}[t]
\centering
\resizebox{1.02 \textwidth}{!}{
 \includegraphics{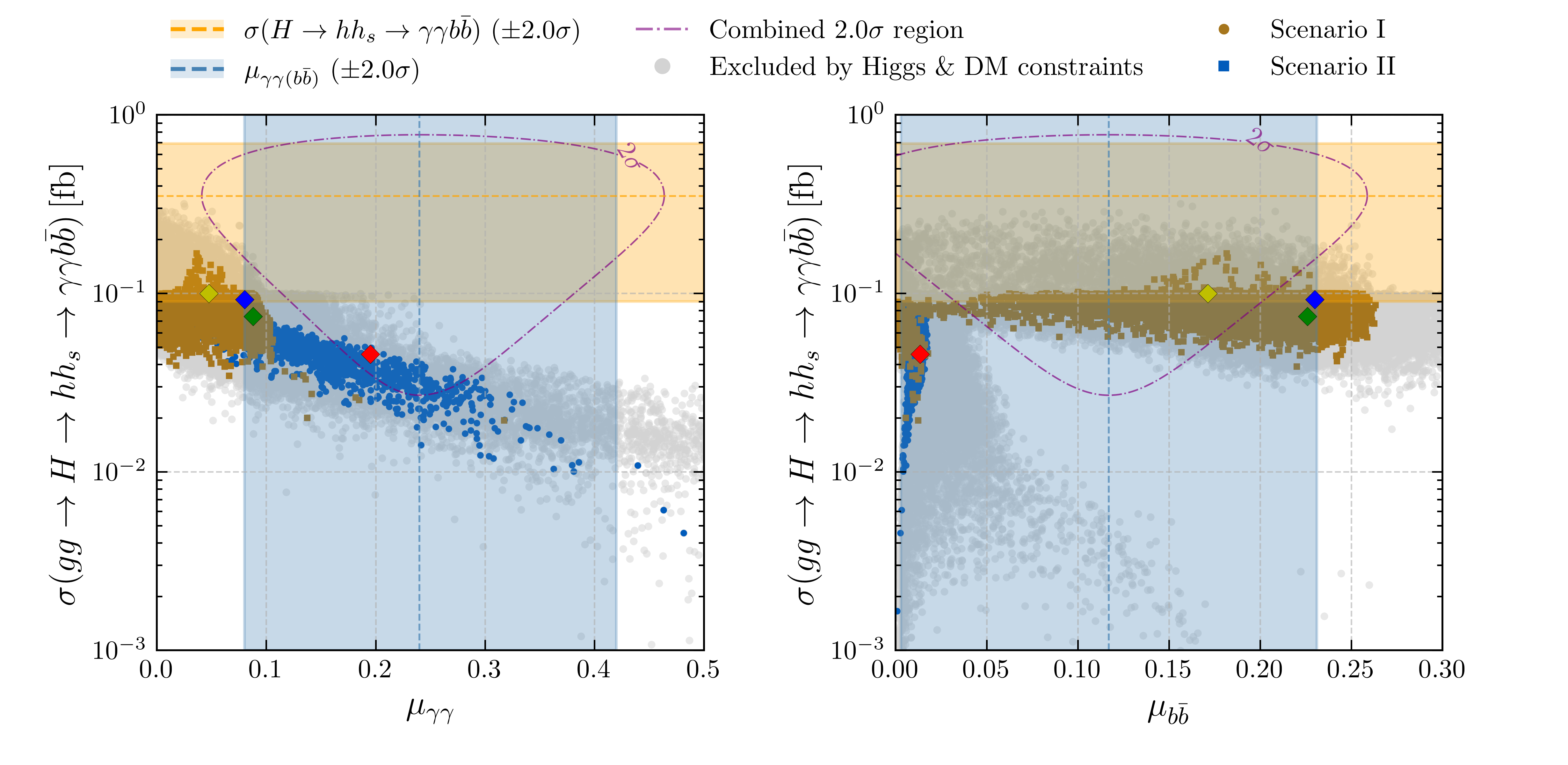}}
\vspace{-0.8cm}
\caption{Scattering plots of samples satisfying all applied constraints, projected onto the planes of the cross section $\sigma(gg\to H\to hh_s\to \gamma\gamma b\bar{b})$ versus the signal strengths $\mu_{\gamma\gamma}$ (left) and $\mu_{b\bar{b}}$ (right). The orange dashed lines and shaded bands indicate the central value of $\sigma(gg\to H\to hh_s\to \gamma\gamma b\bar{b})$ in Eq.~(\ref{XSbbrrExp}) with the corresponding $2\,\sigma$ uncertainty, while the blue lines show the central values of $\mu_{\gamma\gamma}$ in Eq.~(\ref{diphoton-rate}) (left) and $\mu_{b\bar{b}}$ in Eq.~(\ref{LEPrate}) (right). The purple dash-dotted contours mark the combined $2\,\sigma$ regions on each plane. Grey dots represent samples excluded by \textsf{HiggsSignals}, \textsf{HiggsBounds}, or by DM relic density and direct detection constraints. Amber squares and blue dots denote {\bf Scenario I} and {\bf Scenario II}, respectively. Colored diamonds correspond to the four benchmark points, with details listed in Table.~\ref{BP1BP2} and Table.~\ref{BP3BP4}.
\label{Fig1}}
\end{figure}

The correlation between the cross section $\sigma(gg\to H\to hh_s\to \gamma\gamma b\bar{b})$ and the signal strengths $\mu_{\gamma\gamma}$ and $\mu_{b\bar{b}}$ are depicted in the left and right panels in Fig.~\ref{Fig1}, respectively.
 The orange dashed lines and shaded bands mark the central value of the cross section $\sigma_{\gamma\gamma b\bar{b}}$ with the corresponding $2\,\sigma$ uncertainty, and the blue ones mark that of the signal strengths $\mu_{\gamma\gamma}$ and $\mu_{b\bar{b}}$. The purple dashed-dotted contour delineate the combined $2\,\sigma$ region on each plane. 
The plot shows that {\bf Scenario I} can interpret the $\sigma_{\gamma\gamma b\bar{b}}$ excess near 650 GeV and {{$\mu_{\gamma\gamma}$}} and $\mu_{b\bar{b}}$ excesses near 95 GeV at the $2\,\sigma$ level at the same time. 
While samples in {\bf Scenario I}  can push the cross section $\sigma(gg\to H\to hh_s\to \gamma\gamma b\bar{b})$ value closer to the $1\,\sigma$ range from below, this comes at the cost of shifting {  {$\mu_{\gamma\gamma}$}} away from the $2\,\sigma$ range. 
Moreover, these points tend to be excluded by the CMS upper 95\% CL limit of approximately $0.1\rm \;fb$ from the $\tau\bar{\tau}$ plus $b\bar{b}$ search~\cite{CMS:2021yci, Ellwanger:2023zjc}. 
In contrast, $\sigma(gg\to H\to hh_s\to \gamma\gamma b\bar{b})$ in {\bf Scenario II} can reach at most {  { $0.08\rm \;fb$}}.
Nevertheless, it achieves an overall $2\,\sigma$ fit when combining the $\sigma_{\gamma\gamma b\bar{b}}$ and  {  {$\mu_{\gamma\gamma}$}} excess, i.e., $\chi^2_{\sigma_{\gamma\gamma b\bar{b}}} + \chi^2_{\color{blue} {\mu_{\gamma\gamma}}} \leq 6.18$. 
In this scenario, {  {$\mu_{\gamma\gamma}$}} can reach its central value, whereas $\mu_{b\bar{b}}$ remains below $0.02$,  though still  within the $2\,\sigma$ region. 
This behavior arises because a suppression of ${\rm Br}_{\rm SUSY}(h_s \to b \bar{b})$ is needed to enhance the branching ratio ${\rm Br}_{\rm SUSY}(h_s \to \gamma\gamma)$ and thus a larger {  {$\mu_{\gamma\gamma}$}}. 
This also explains the trend seen in the left panel in Fig.~\ref{Fig1}, where $\sigma(gg\to H\to hh_s\to \gamma\gamma b\bar{b})$ decreases with {  {$\mu_{\gamma\gamma}$}} increases.
Samples excluded by constraints from the Higgs experiments, DM relic density and direct detections are denoted by grey dots. In particular, the searches for additional Higgs states implemented in \textsf{HiggsBounds} set the strongest limits for $\sigma(gg\to H\to hh_s\to \gamma\gamma b\bar{b}) \gtrsim 0.09 \rm \;fb$, while the SM-like Higgs precision data encoded in \textsf{HiggsSignals} dominate the exclusions below this threshold.  Furthermore, roughly half of the remaining samples are removed by DM relic density and direct detection constraints.

\begin{figure}[t]
\centering
\resizebox{1.02 \textwidth}{!}{
 \includegraphics{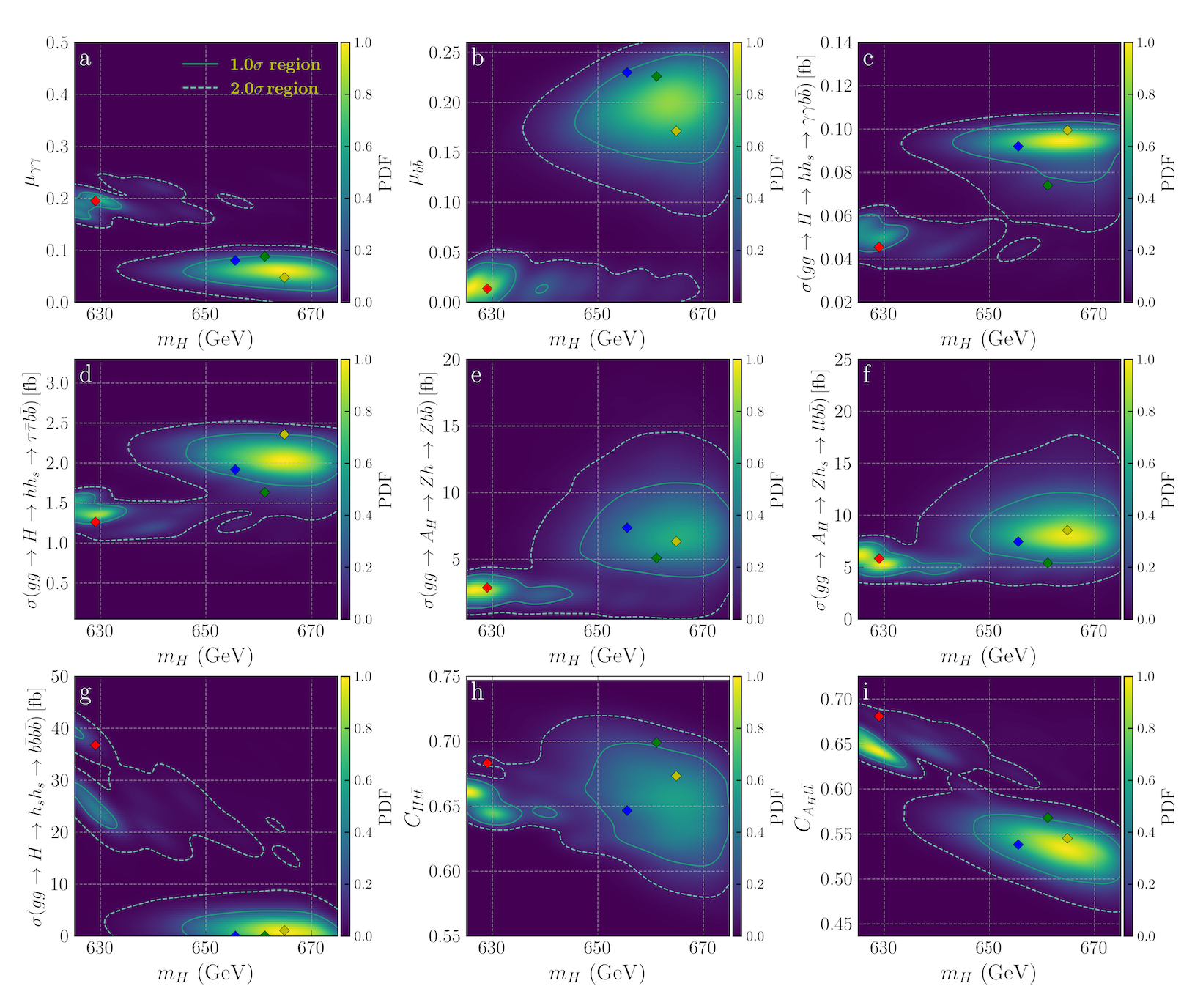}}
\vspace{-0.8cm}
\caption{ 
Two-dimensional posterior PDF distributions of the heavy Higgs boson mass $m_H$ versus (a) $\mu_{\gamma\gamma}$, (b) $\mu_{b\bar{b}}$, (c) $\sigma(gg\to H\to hh_s\to \gamma\gamma b\bar{b})$, (d) $\sigma(gg\to H\to hh_s\to \tau\bar{\tau}b\bar{b})$, (e) $\sigma(gg\to A_H\to Zh \to Zb\bar{b})$, (f) $\sigma(gg\to A_H\to Zh_s \to \ell\ell b\bar{b})$, (g) $\sigma(gg\to H\to h_sh_s\to b\bar{b}b\bar{b})$, (h) $C_{Ht\bar{t}}$, and (i) $C_{A_Ht\bar{t}}$. Solid and dashed contours indicate the $1\,\sigma$ and $2\,\sigma$ credible regions, respectively. Colored diamonds denote the four benchmark points corresponding to those in Fig.~\ref{Fig1}, with details provided in Table.~\ref{BP1BP2} and Table.~\ref{BP3BP4}.
\label{Fig2}}
\end{figure}

The right panel of Fig.\ref{Fig1} displays a nontrivial correlation between $\sigma(gg\to H\to hh_s\to \gamma\gamma b\bar{b})$ and $\mu_{b\bar{b}}$. This complexity arises because an increase ${\rm Br}_{\rm SUSY}(h_s \to b \bar{b})$ does not necessarily translate into a corresponding enhancement of $\mu_{b\bar{b}}$. In particular, increasing the coupling of $h_s$ to bottom quark pairs $|C_{h_s b \bar{b}}|$ may simultaneously suppress its couplings to the electroweak gauge bosons $|C_{h_s V V}|$~\footnote{Recalling the normalized couplings of $h_i$ to $b \bar{b}$ and vector bosons  $V=W,Z$ at tree level in terms of the mixing coefficients $V^j_{h_i}$ in Eq.(\ref{Vij}) :
\begin{eqnarray}
C_{h_i b \bar{b}} =  V_{h_i}^{\rm SM} - V_{h_i}^{\rm NSM} \tan \beta,  \quad C_{h_i V V} = V_{h_i}^{\rm SM},  \nonumber
\label{hs-couplings}
\end{eqnarray}
$C_{h_s V V}$ may decrease with $C_{h_s b \bar{b}}$ increases when $V_{h_s}^{\rm SM}$ and $V_{h_s}^{\rm NSM}$ share the same sign. In this study, we verify that this behaviour occurs for $ 0.0 \lesssim C_{h_s b \bar{b}} \lesssim 0.3$. 
Furthermore, the effective coupling of the SM-like Higgs $h$ to $\gamma\gamma$ decreases from about 1.0 to 0.89 once $C_{h_s b \bar{b}} \gtrsim 0.3$.}. 
Because $\mu_{b\bar{b}}$ depends not only on ${\rm Br}_{\rm SUSY}(h_s\to b\bar{b})$ but also on the associated suppression of the $h_s$--$Z$ coupling, $\mu_{b\bar{b}}$ may saturate or even decrease as ${\rm Br}_{\rm SUSY}(h_s\to b\bar{b})$ grows.
Although in most of the parameter space $|C_{h_s b\bar{b}}|$ and $|C_{h_s V V}|$ increase together, the associated reduction of the SM-like Higgs coupling to photons suppresses ${\rm Br}_{\rm SUSY}(h \to \gamma\gamma)$. This effect offsets the enhancement of $\sigma(gg\to H\to hh_s\to \gamma\gamma b\bar{b})$ expected from increasing ${\rm Br}_{\rm SUSY}(h_s\to b\bar{b})$, thereby further contributing to the intricate structure in the $\mu_{b\bar{b}}$–$\sigma(gg\to H\to hh_s\to \gamma\gamma b\bar{b})$ correlation.

Next, we apply the Bayesian inference method to analyze the model predictions for the heavy Higgs boson $H$ near 650 GeV.  By accumulating the posterior PDF of samples obtained from the scan process~\footnote{The nested sampling algorithm used in this work transforms the multidimensional evidence integral into a one-dimensional integral while sampling the parameter space~\cite{Importance2019,MultiNest2009,Feroz:2007kg,Skilling:2006gxv}. The evidence integral can be written as~\cite{MultiNest2009}:
\begin{equation}
\mathcal{Z} = \int^1_0 \mathcal{L}(X)dX, \quad   X(\lambda) = \int_{\mathcal{L}(\Theta)} \pi(\Theta) d^D\Theta, \nonumber
\label{BayesianEv}
\end{equation}
where $\Theta$ is the input parameters of the model, $D$ the dimensionality of the parameter space, $\pi(\Theta)$ the prior, $\mathcal{L}(\Theta)$ the likelihood, and $X$ the ‘prior volume’ with the definition $dX=\pi(\Theta) d^D\Theta$. }, 
we find that {\bf Scenario I} and {\bf Scenario II} contribute approximately $68\%$ and $23\%$ of the total Bayesian evidence, respectively, indicating a clear preference of current experimental data for the former.  
The posterior PDF distributions of $\mu_{\gamma\gamma}$, $\mu_{b\bar{b}}$, and $\sigma(gg\to H\to hh_s\to \gamma\gamma b\bar{b})$ as functions of the heavy Higgs mass $m_H$ are shown in the upper panels of Fig.~\ref{Fig2}.
The solid contours indicate the $1\,\sigma$ credible regions. For $625~{\rm GeV}\lesssim m_H \lesssim 640~{\rm GeV}$, we find:
\begin{eqnarray}
0.16 \lesssim \mu_{\gamma\gamma} \lesssim 0.21,\quad 0.0 \lesssim \mu_{b\bar{b}} \lesssim 0.04, \quad 0.04 {\rm \;fb} \lesssim \sigma(gg\to H\to hh_s\to \gamma\gamma b\bar{b})  \lesssim 0.06 {\rm \;fb}.\nonumber
\end{eqnarray}
For $640 {\rm \; GeV}\lesssim m_H \lesssim 675 {\rm \; GeV}$, we find:
\begin{eqnarray}
0.02 \lesssim \mu_{\gamma\gamma} \lesssim 0.09, \quad 0.14 \lesssim \mu_{b\bar{b}} \lesssim 0.25, \quad 0.07 {\rm \;fb} \lesssim \sigma(gg\to H\to hh_s\to \gamma\gamma b\bar{b})  \lesssim 0.105 {\rm \;fb}.\nonumber
\end{eqnarray}
As expected, {\bf Scenario~I} and {\bf Scenario~II} primarily populate the regions with $m_H \lesssim 640~{\rm GeV}$ and $m_H \gtrsim 640~{\rm GeV}$, respectively. 
The pattern displayed in the $\mu_{\gamma\gamma}-m_H$ and $\mu_{b\bar{b}}-m_H$ planes can be understood from the fact that heavier mass of $m_H$ requires an increasing ${\rm Br}_{\rm SUSY}(h_s \to b \bar{b})$ to keep $\sigma(gg\to H\to hh_s\to \gamma\gamma b\bar{b})$ from decreasing for kinematic reasons, thus leading to a higher $\mu_{b\bar{b}}$ and lower $\mu_{\gamma\gamma}$.   

As mentioned earlier, the cross section $\sigma(gg\to H\to hh_s\to \tau\bar{\tau}b\bar{b})$ is constrained to be below about $3~{\rm fb}$ for $600~{\rm GeV}\lesssim m_H \lesssim 700~{\rm GeV}$ by CMS~\cite{CMS:2021yci}. Panel~(d) of Fig.~\ref{Fig2} shows that both the $1\,\sigma$ and $2\,\sigma$ credible regions lie well within the current experimental bounds.
 
We further explore several phenomenologically interesting search channels and present their prospects as tests of the model. The scenarios studied here predict a doublet–like CP–odd Higgs boson $A_H$ with a mass range of $450~{\rm GeV}\lesssim m_{A_H} \lesssim 650~{\rm GeV}$ (corresponding to $m_H$ varying from $\sim 675~{\rm GeV}$ to $\sim 625~{\rm GeV}$) at the $1\,\sigma$ level. 
Searches in the channel $\sigma(gg\to A_H\to Zh \to Zb\bar{b})$ conducted by CMS~\cite{CMS:2019qcx} and ATLAS~\cite{ATLAS:2022enb} place a $95\%$~CL upper limit of roughly $30~{\rm fb}$ near $m_{A_H}\simeq 600~{\rm GeV}$ (or equivalently $m_H\simeq 645~{\rm GeV}$). 
This limit relaxes to about $100~{\rm fb}$ as $m_{A_H}$ decreases to $450~{\rm GeV}$ (or $m_H$ increases to $675~{\rm GeV}$). Searches in the complementary channel $\sigma(gg\to A_H\to Zh_s\to \ell\ell b\bar{b})$ by CMS~\cite{CMS:2019ogx} impose upper limits between $20$ and $30~{\rm fb}$ for $625~{\rm GeV}\lesssim m_H \lesssim 675~{\rm GeV}$. 
Panels (e) and (f) of Fig.~\ref{Fig2} display the corresponding predictions, demonstrating that they lie safely below the current bounds and are likely to be probed at future high–luminosity collider runs.

\begin{figure}[t]
\centering
\resizebox{0.86 \textwidth}{!}{
 \includegraphics{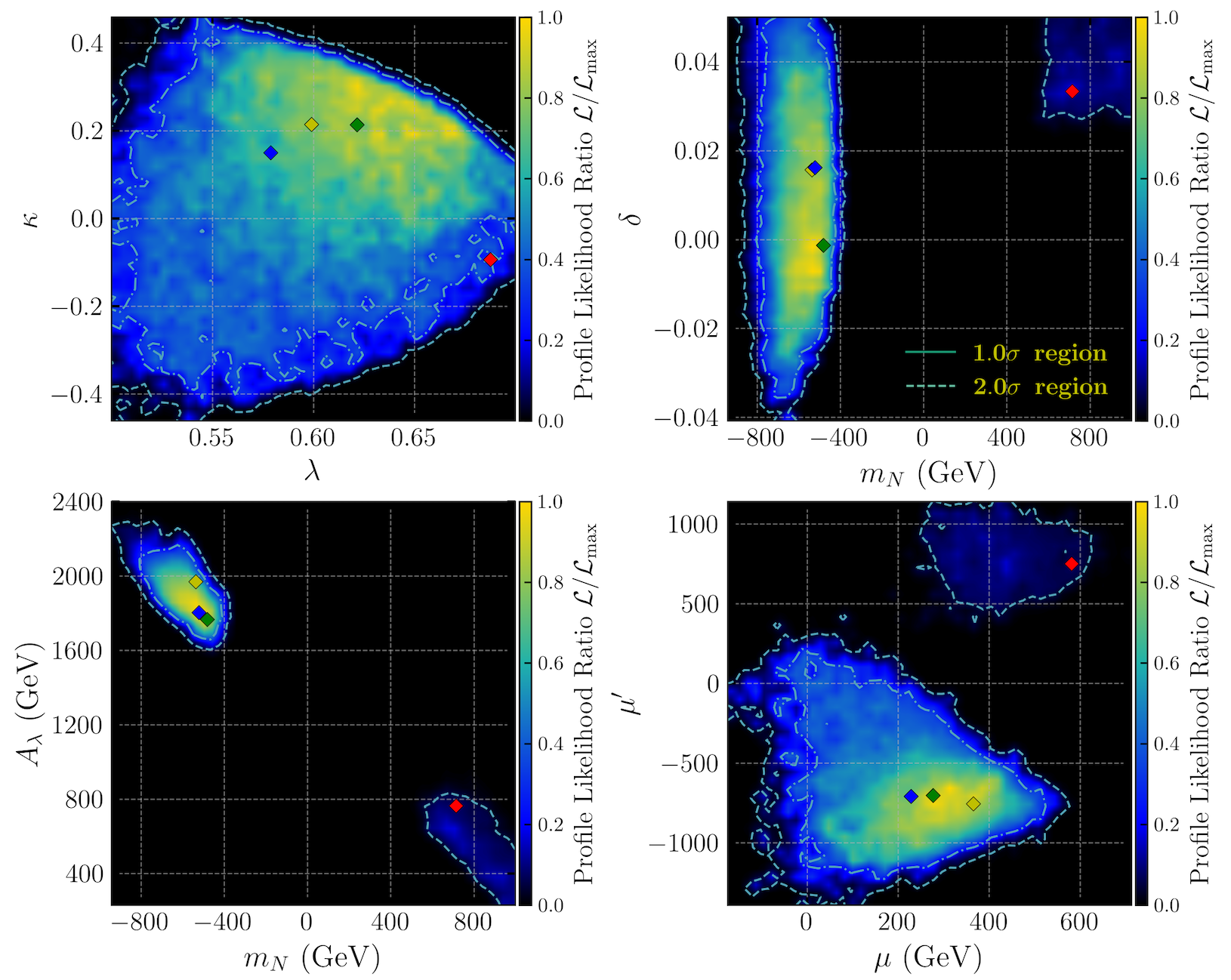}}
\vspace{-0.2cm}
\caption{Two-dimensional profile likelihoods projected onto the $(\lambda, \kappa)$, $(m_N, \delta)$, $(m_N, A_\lambda)$, and $(\mu, \mu^\prime)$ planes. Dashed-dotted and dashed contours indicate the $1\,\sigma$ and $2\,\sigma$ confidence intervals, respectively. The best-fit point has a total $\chi^2$ value of $\chi^2_{650+95} + \chi^2_{125} \simeq 162$. Colored diamonds denote the four benchmark points shown in Figs.~\ref{Fig1} and \ref{Fig2}.
\label{Fig3}}
\end{figure}

Considering that the decays of $H$ or $A_H$ into top–quark pairs may provide promising search channels, we show the normalized couplings $C_{H t\bar{t}}$ and $C_{A_H t\bar{t}}$ in panels (h) and (i) of Fig.~\ref{Fig2}, respectively. 
The CMS search for heavy resonances decaying into $t\bar{t}$~\cite{CMS:2019pzc} sets an approximate upper limit of $C_{Ht\bar{t}} \simeq 0.86$ for the ratio of width to mass $\Gamma_H/m_H \simeq 2.5\%$, and $C_{A_Ht\bar{t}} \simeq 0.75$ for $\Gamma_{A_H}/m_{A_H} \simeq 2.5\%$ with $m_{H} \simeq 629 {\rm \; GeV}$. 
It is worth noting that in panel (i) of Fig.~\ref{Fig2} the region with $C_{A_H t\bar{t}} \lesssim 0.60$ corresponds entirely to {\bf Scenario~I}, while values above $0.6$ correspond to {\bf Scenario~II}.
The origin of this separation is clarified later in the discussion of the lower–left panel of Fig.~\ref{Fig3}.

We also calculate the cross sections of H decaying into the SM-like Higgs pairs $\sigma(gg\to H \to hh)$ and the light Higgs pairs $\sigma(gg\to H \to h_sh_s)$. 
The largest value in the $1\,\sigma$ region of $\sigma(gg\to H \to hh)$ is about $20 \rm \;fb$ for {\bf Scenario I} and $3.5 \rm \;fb$ for {\bf Scenario II}. 
For the $h_sh_s$ channel, this channel is significantly more promising, with the largest cross-section reaching $25 \text{ fb}$ for {\bf Scenario~I} and a substantial $120 \text{ fb}$ for {\bf Scenario~II}. 
The largest predicted values of $\sigma(gg\to H \to h_s h_s)$ multiplied by the relevant branching fractions into the final states $b\bar{b}b\bar{b}$, $b\bar{b}\tau\bar{\tau}$, and $b\bar{b}\gamma\gamma$ are approximately $36 \rm \;fb$, $4.5 \rm \;fb$ and $0.23 \rm \;fb$, respectively. 
The distribution of the $\sigma(gg\to H\to h_sh_s\to b\bar{b}b\bar{b})$ cross-section is displayed in panel (g) of Fig.~\ref{Fig2}.

Fig.~\ref{Fig3} shows the two-dimensional PL functions projected onto the $(\lambda, \kappa)$, $(m_N,\delta)$, $(m_N, A_\lambda)$ and $(\mu, \mu^\prime)$ planes~\footnote{The two-dimensional profile likelihood for parameters 
$(\theta_1,\theta_2)$ is defined as
\begin{equation}
\mathcal{L}_{\rm prof}(\theta_1,\theta_2)= \max_{\theta_3,\ldots,\theta_n}\,\mathcal{L}(\theta_1,\theta_2,\theta_3,\ldots,\theta_n).  \nonumber
\end{equation}
At each point in the $(\theta_1,\theta_2)$ plane, it represents the maximum likelihood compatible with the data after optimizing over all remaining nuisance parameters. The profile likelihood therefore encodes the maximally allowed parameter region consistent with current experimental data}. 
The dashed-dotted and dashed contours delineate the $1\,\sigma$ and $2\,\sigma$ confidence intervals, respectively. 
The best-fit point yields a combined value of $\chi^2_{650+95} + \chi^2_{125} \simeq 162$. 
The upper-left panel shows that the $1\,\sigma$ regions of $\lambda$ and $\kappa$ span nearly the entire prior range, bounded only by the perturbativity condition. The preferred region lies at relatively large and positive values of $(\lambda,\kappa)$, where {\bf Scenario~I} and {\bf Scenario~II} overlap.  
In contrast, the remaining three panels reveal two clearly separated islands of viable solutions. We verify that
 \begin{eqnarray}
{\bf Scenario~I}:\, m_N \lesssim -390 \,\rm GeV, \quad A_\lambda \gtrsim 1600 \,\rm GeV, \quad \mu^\prime \lesssim 400 \,\rm GeV,\nonumber
\end{eqnarray}
while 
\begin{eqnarray}
{\bf Scenario~II}: \, m_N \gtrsim 550 \,\rm GeV, \quad A_\lambda \lesssim 800 \,\rm GeV, \quad \mu^\prime \gtrsim 500 \,\rm GeV, \nonumber
\end{eqnarray}
corresponding only to the $2\,\sigma$ credible region.

The structure seen in the lower-left panel of Fig.~\ref{Fig3} indicates that $A_\lambda + m_N$ is confined to nearly the same range in both scenarios. 
This behavior follows from the requirement that $A_\lambda + m_N \simeq 2\mu_{tot}$ (given $\tan\beta \simeq 1.6$), which is necessary to keep $\delta$ in the small range demanded by Eq.~(\ref{delta}). 
Moreover, in {\bf Scenario~I}, the enhancement of $A_\lambda - m_N$ leads to a relatively larger mixing between $A_{\rm NSM}$ and $A_S$ [see Eq.~(\ref{CPoddHiggsMass})], thereby suppressing the coupling of $A_H$ to top-quark pairs.
Conversely, in {\bf Scenario~II}, the partial cancellation between $A_\lambda$ and $m_N$ yields a larger $C_{A_H t\bar{t}}$, consistent with the pattern shown in panel (i) of Fig.~\ref{Fig2}.
Finally, the $(\mu,\mu^\prime)$ panel reveals $1\,\sigma$ and $2\,\sigma$ regions that significantly deviate from zero, suggesting that non-vanishing values of $\mu$ and $\mu^\prime$ may serve as distinguishing features of the GNMSSM relative to the $\mathbb{Z}_3$-NMSSM and the MSSM.

\begin{figure}[t]
\centering
\resizebox{1.06 \textwidth}{!}{
 \includegraphics{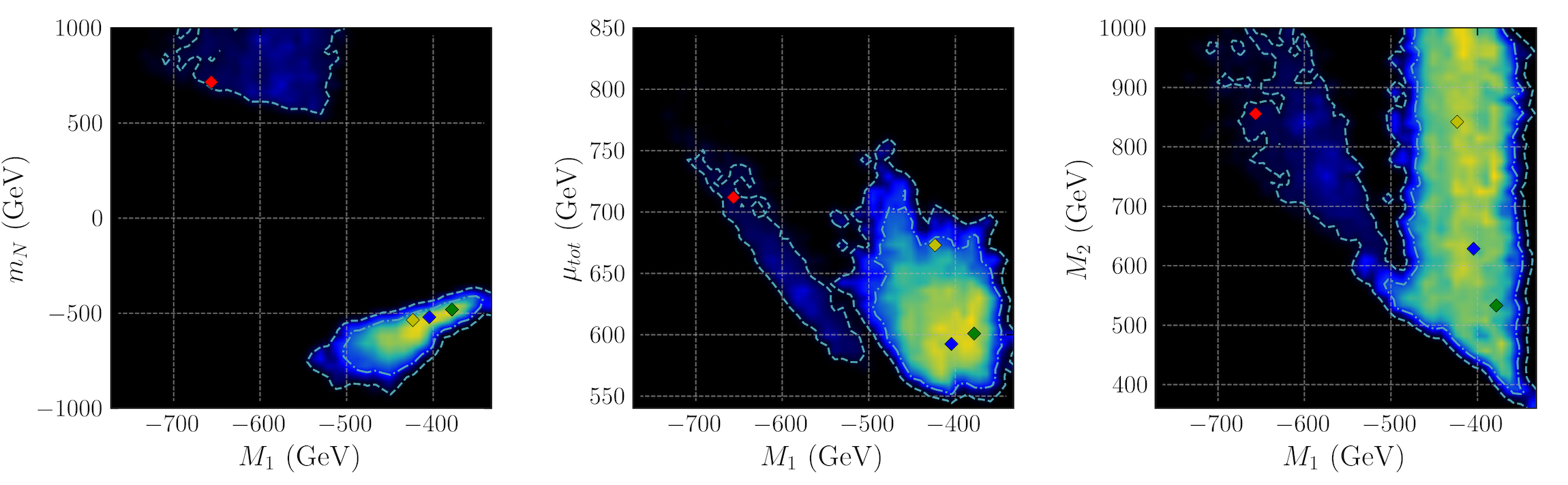}}
\vspace{-0.6cm}
\caption{Same as Fig.~\ref{Fig3}, but projected onto $(M_1, m_N)$, $(M_1, \mu_{tot})$, $(M_1, M_2)$ planes. \label{Fig4}}
\end{figure}

Lastly, we discuss the DM dynamics relevant for explaining the observed excesses. 
Since both scenarios predict a Bino-dominated neutralino $\tilde{\chi}^0_1$ as the DM candidate, we show in Fig.~\ref{Fig4} the PL functions of the Bino mass parameter $M_1$ correlated with the Singlino mass $m_N$, the total Higgsino mass parameter $\mu_{\rm tot}$, and the Wino mass $M_2$.
From the left panel of Fig.~\ref{Fig4}, one observes that the $1\,\sigma$ region of $M_1$ in {\bf Scenario~I} lies in the range $-500 {\, \rm GeV} \lesssim M_1 \lesssim -300 \, \rm GeV$. 
In this scenario, the relic density is obtained either through efficient $s$-channel resonance annihilation $\tilde{\chi}^0_1\tilde{\chi}^0_1 \to h_sA_H$, which is enhanced when $2m_{\tilde{\chi}^0_1} \simeq m_{A_s} \simeq 800 \, \rm GeV$, or through coannihilation with Singlino-dominated $\tilde{\chi}^0_2$ states.
In the latter case, the dominant processes are $\tilde{\chi}^0_2\tilde{\chi}^0_2, \tilde{\chi}^0_1\tilde{\chi}^0_2 \to h_sA_H$,
which together yield the correct DM abundance.
In contrast, in {\bf Scenario~II}, the $2\,\sigma$ region favors heavier and more negative $M_1$ values $M_1 \lesssim -500 \, \rm GeV$.  
As seen from the middle panel of Fig.~\ref{Fig4}, the relic density is predominantly controlled by coannihilation with Higgsino-dominated $\tilde{\chi}^0_2$ or $\tilde{\chi}^\pm_1$ states. The leading channels in this case are $\tilde{\chi}^+_1\tilde{\chi}^-_1, \tilde{\chi}^0_2\tilde{\chi}^0_2 \to h_sA_s$, as well as processes such as $\tilde{\chi}^+_1\tilde{\chi}^0_2  \to u\bar{d}, h_sW^+$. These channels efficiently reduce the relic abundance to the observed level.

\begin{figure}[t]
\centering
\resizebox{1.05 \textwidth}{!}{
 \includegraphics{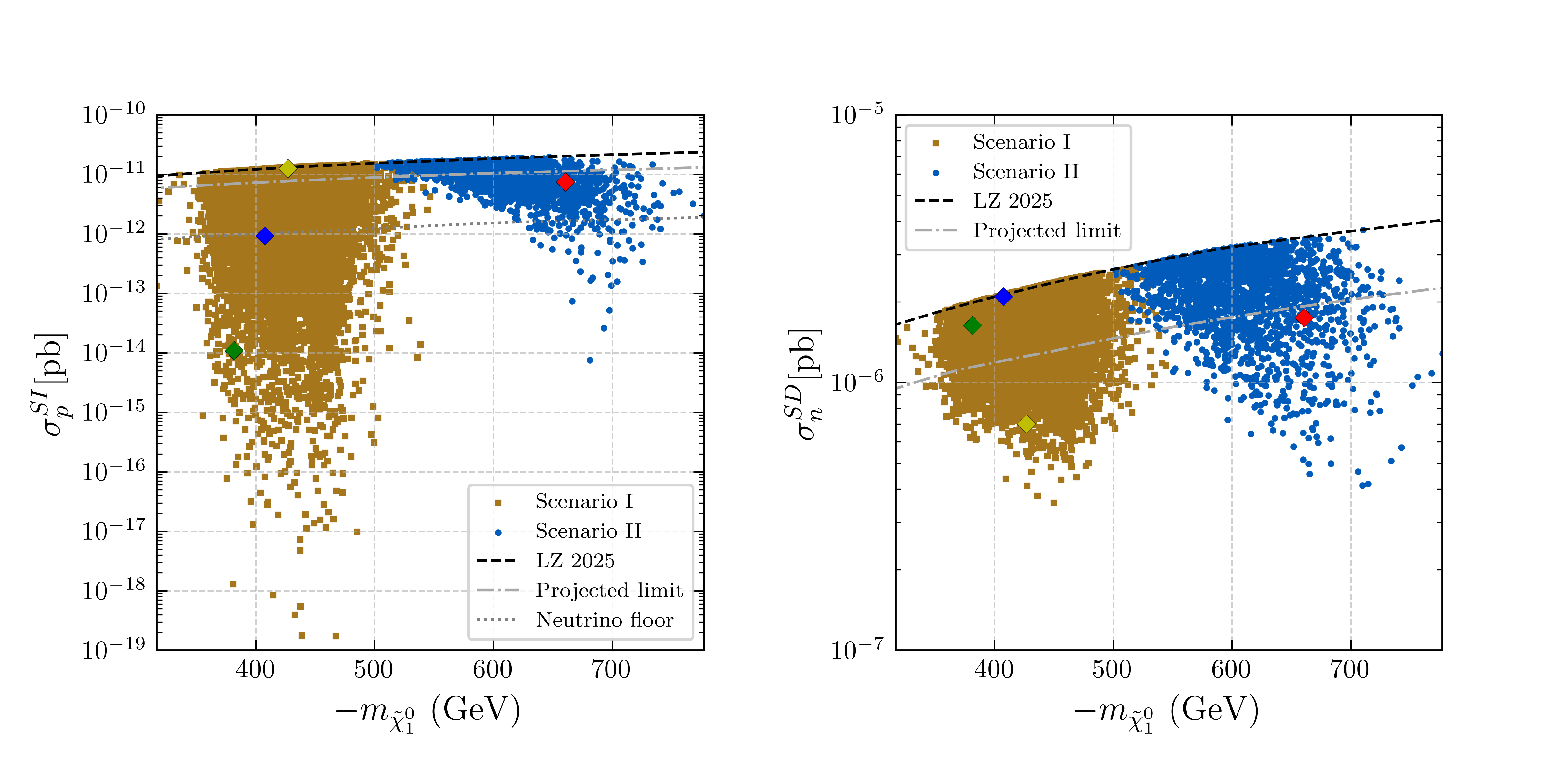}}
\vspace{-0.6cm}
\caption{ Spin-independent (SI) and spin-dependent (SD) DM–nucleon scattering cross sections, $\sigma^{\rm SI}_p$ (left) and $\sigma^{\rm SD}_n$ (right), as functions of the LSP mass $m_{\tilde{\chi}_1^0}$. Dashed black lines show the 90\% C.L. upper limits from the 2025 LZ results~\cite{LZ:2024zvo,LZ2024slides}, dashed-dotted grey lines indicate the projected LZ sensitivities for future runs~\cite{LZ:2018qzl}, and dotted lines denote the neutrino floor~\cite{Billard:2013qya}. Amber squares and blue dots correspond to {\bf Scenario I} and {\bf Scenario II}, respectively. Colored diamonds represent the four benchmark points shown in Figs.~\ref{Fig1}–\ref{Fig4}.
 \label{Fig5}}
\end{figure}

In Fig.~\ref{Fig5}, the SI and SD DM–nucleon scattering cross sections are shown as functions of the LSP mass $m_{\tilde{\chi}_1^0}$ in the left and right panels, respectively. 
For comparison, we overlay the latest $90\%$~CL upper limits from the combined WS2024+WS2022 analysis of the LZ experiment~\cite{LZ:2024zvo, LZ2024slides}, the projected sensitivities for future LZ runs~\cite{LZ:2018qzl}, and the expected neutrino background~\cite{Billard:2013qya}. 
The left panel illustrates that most viable samples lie well below both the current and projected SI limits.
Within {\bf Scenario~I}, samples with $\sigma(gg\to H\to hh_s\to \gamma\gamma b\bar{b}) > 0.09~{\rm fb}$ in the lower $m_{\tilde{\chi}1^0}$ region generally predict smaller $\sigma^{\rm SI}_p$ compared with {\bf Scenario~II}, whose points satisfying $\chi^2_{\sigma_{\gamma\gamma b\bar{b}}}+\chi^2_{{\color{blue} {\mu_{\gamma\gamma}}}} \leq 6.18$ appear at larger $m_{\tilde{\chi}_1^0}$.
The right panel shows that while all samples remain consistent with current SD bounds, the projected sensitivity can probe roughly half of them.

To further clarify the model’s interpretation of the observed excesses, four benchmark points—P1, P2, and P3 from {\bf Scenario~I}, and P4 from {\bf Scenario~II}—are selected to illustrate representative behaviors. These benchmark points are highlighted as blue, yellow, green, and red diamonds, respectively, in all relevant figures, and their detailed properties are listed in Tables.~\ref{BP1BP2} and~\ref{BP3BP4}.
{  {The phenomenological characteristics of these benchmarks are summarized as follows:
\begin{itemize}
\item
P1 simultaneously accommodates the $\sigma_{\gamma\gamma b\bar{b}}$ excess near 650 GeV, as well as the excesses in $\mu_{\gamma\gamma}$ and $\mu_{b\bar{b}}$ near 95~GeV,  all at the $2\,\sigma$ level (treated individually rather than in combination).
\item
P2 provides a $2\,\sigma$ interpretation of the combined excesses in $\sigma_{\gamma\gamma b\bar{b}}$ and $\mu_{b\bar{b}}$, and predicts $\mu_{\gamma\gamma}$ within $2.4\,\sigma$. 
\item
P3 successfully explains the $\mu_{\gamma\gamma}$ and $\mu_{b\bar{b}}$ excesses at the $2\,\sigma$ level, while slightly exceeding the $2\,\sigma$ preferred region of $\sigma_{\gamma\gamma b\bar{b}}$.
\item
P4 yields a $2\,\sigma$ interpretation of the combined $\sigma_{\gamma\gamma b\bar{b}}$ and $\mu_{\gamma\gamma}$ anomalies, with $\mu_{b\bar{b}}$ falling within the $2\,\sigma$ interval as well.
 \end{itemize}
Concerning DM physics, the $\tilde{\chi}^0_1$ candidate in P1, P2 and P3 achieves the observed relic abundance primarily via the $A_s$-funnel mechanism, annihilating into the $h_s A_H$ final state. In particular, P1 and P3 rely on coannihilation processes involving a $\tilde{S}$-like $\tilde{\chi}_2^0$, whereas P4 requires coannihilation with $\tilde{H}$-like electroweakinos. 
In terms of direct detection prospects, P1 and P3 may be probed by future SD searches. However, their SI scattering cross sections are strongly suppressed and lie below the neutrino floor, due to significant cancellations among different contributions to the $\tilde{\chi}^0_1$-Higgs boson coupling.
In contrast, P2 predicts an extremely small SD neutron scattering cross section, as low as $6.9\times 10^{-7} \,\rm pb$, resulting from a highly suppressed $\tilde{\chi}^0_1$-Z gauge boson coupling characterized by $N_{13}^2-N_{14}^2 \simeq 0.004$. }}

Regarding constraints from the LHC searches for SUSY, simulation results of electroweakino productions, utilizing the package $\texttt{CheckMATE-2.0.26}$, consistently indicate that the constraints from LHC searches for new physics are minimal. The calculated $R$-values for all benchmark points are less than $0.35$, confirming that the parameter space is safely compliant with existing $95\%$ CL limits.
The basic reason for these weak restrictions is that the interpretation of the scalar excesses primarily relies on Higgs sector mixing effects rather than sizable contributions from loop effects involving light SUSY particles.
Specific analysis of the benchmark points highlights the dominant search channels:
\begin{itemize}
\item {  { P1, P2 and P3: The most significant signal region is G05 from the CMS report $\texttt{CMS-SUS-16-039}$ \cite{CMS:2017moi}. This targets events with at least four leptons, no $\tau$ leptons, and high missing transverse momentum ($p^{\rm miss}_T \geq 200 \text{ GeV}$).}}
\item P4: This point yields the lowest $R$ value because it predicts a highly compressed mass spectrum, with a mass difference $\Delta m (\tilde{\chi}^0_2,\tilde{\chi}^0_1) \lesssim 20 \text{ GeV}$, making the decay products too soft to be effectively detected by current searches.
 \end{itemize}
For each benchmark, the most significant signal regions, corresponding $R$ values, and the associated experimental analyses are precisely listed in the bottom rows of Table.~\ref{BP1BP2} and Table.~\ref{BP3BP4}.

\begin{table}[t]
%\begin{sidewaystable}
\centering
\caption{\label{BP1BP2} Details of the benchmark points P1 and P2 from {\bf Scenario I}, represented as blue and yellow diamonds, respectively, in all figures. {  { P1 simultaneously accommodates the $\sigma_{\gamma\gamma b\bar{b}}$ excess near 650 GeV, as well as the excesses in $\mu_{\gamma\gamma}$ and $\mu_{b\bar{b}}$ near 95~GeV at the $2\,\sigma$ level, while satisfying all applied constraints. P2 provides a $2\,\sigma$ combined interpretation of the $\sigma_{\gamma\gamma b\bar{b}}$ and $\mu_{b\bar{b}}$ excesses.}} }
\vspace{0.3cm}
\resizebox{1\textwidth}{!}
 {
\begin{tabular}{lrlr|lrlr}
\hline \hline
\multicolumn{4}{c|}{\bf Benchmark Point P1 }                                                                                                				& \multicolumn{4}{c}{\bf Benchmark Point P2}
 \\ \hline
$\mu_{tot}$                		& 592.4~GeV 	&$\lambda$             		& 0.579				&$\mu_{tot}$                & 672.9~GeV 	&$\lambda$             		& 0.599 \\
$\mu$					& 228.7~GeV	&$\kappa$              		& 0.150				&$\mu$				& 365.6~GeV	&$\kappa$              		& 0.214\\
$\mu^\prime$			& -709.1~GeV	&$\delta$         			& 0.016				&$\mu^\prime$		& -755.8~GeV	&$\delta$          			& 0.016\\
$M_1$               			& -404.3~GeV 	&$\tan{\beta}$                	& 1.557				&$M_1$               		& -423.3~GeV 	&$\tan{\beta}$                	& 1.493\\
$M_2$             		 	& 628.3~GeV	&$A_{\lambda}$	 		&1803~GeV			&$M_2$             		& 842.0~GeV	&$A_{\lambda}$ 		       & 1969~GeV	\\
$m_A$					& 603.7~GeV	&$A_{\kappa}$			&1808~GeV			&$m_A$				&626.4~GeV		&$A_{\kappa}$			& 1893~GeV  \\
$m_B$					& 113.4~GeV	&$A_t$					& 2005~GeV			&$m_B$				&118.5~GeV		&$A_t$					&1662~GeV	\\
$m_N$					& -521.1~GeV	&						&					&$m_N$				&-536.2~GeV	&						&   \\
\hline
$m_{\tilde{\chi}_1^0}$   	 & -407.0~GeV	&$m_{h_s}$                	& 95.0~GeV			&$m_{\tilde{\chi}_1^0}$   		& -427.2~GeV	&$m_{h_s}$                		& 95.71~GeV\\
$m_{\tilde{\chi}_2^0}$   	 & -451.6~GeV	&$m_{h}$              	& 127.1~GeV		&$m_{\tilde{\chi}_2^0}$   		& -478.4~GeV	&$m_{h}$              		& 124.1~GeV\\
$m_{\tilde{\chi}_3^0}$   	 & 547.0~GeV	&$m_{H}$                  	& 655.6~GeV		&$m_{\tilde{\chi}_3^0}$   		& 654.2~GeV	&$m_{H}$                  		& 664.9~GeV\\
$m_{\tilde{\chi}_4^0}$   	 & -660.0~GeV	&$m_{A_H}$                & 512.1~GeV		&$m_{\tilde{\chi}_4^0}$   		& -728.4~GeV	&$m_{A_H}$                 	& 495.5~GeV\\
$m_{\tilde{\chi}_5^0}$   	 & 707.2~GeV	&$m_{A_s}$               	& 878.2~GeV		&$m_{\tilde{\chi}_5^0}$   		& 896.6~GeV	&$m_{A_s}$                		& 888.6~GeV\\
$m_{\tilde{\chi}_1^\pm}$	 & 505.7~GeV	&$m_{H^\pm}$		& 635.0~GeV		&$m_{\tilde{\chi}_1^\pm}$		& 652.6~GeV	&$m_{H^\pm}$			& 647.6~GeV\\
$m_{\tilde{\chi}_2^\pm}$	 & 706.8~GeV	& 					& 					&$m_{\tilde{\chi}_2^\pm}$	 	& 896.7~GeV	&						& \\
\hline
\multicolumn{2}{l}{$\Omega h^2 $}			&\multicolumn{2}{l|}{0.132}
&\multicolumn{2}{l}{$\Omega h^2 $}		&\multicolumn{2}{l}{0.120}\\
\multicolumn{2}{l}{$\sigma^{SI}_p$,  ~$\sigma^{SD}_n$}		&\multicolumn{2}{l|}{$9.2\times 10^{-49}{\rm ~cm^2}, ~2.0\times 10^{-42}{\rm ~~cm^2}$}
&\multicolumn{2}{l}{$\sigma^{SI}_p$, ~$\sigma^{SD}_n$}		&\multicolumn{2}{l}{$1.2\times 10^{-47}{\rm ~cm^2},  ~6.9\times 10^{-43}{\rm ~~cm^2}$}\\
\hline
\multicolumn{2}{l}{$\mu_{\gamma\gamma}$}		&\multicolumn{2}{l|}{0.081}		
&\multicolumn{2}{l}{$\mu_{\gamma\gamma}$}	&\multicolumn{2}{l}{0.048} \\ 

\multicolumn{2}{l}{$ \mu_{b\bar{b}}$}		&\multicolumn{2}{l|}{0.230}		
&\multicolumn{2}{l}{$\mu_{b\bar{b}}$}		&\multicolumn{2}{l}{0.171} \\ 

\multicolumn{2}{l}{$\sigma_{\gamma\gamma b\bar{b}}$}		&\multicolumn{2}{l|}{$0.092 ~\rm fb$}		
&\multicolumn{2}{l}{$\sigma_{\gamma\gamma b\bar{b}}$}		&\multicolumn{2}{l}{$0.099 ~\rm fb$} \\ 

\multicolumn{2}{l}{$\sigma(gg\to H\to hh_s\to \tau\bar{\tau}b\bar{b}) $}		&\multicolumn{2}{l|}{$1.918 ~\rm fb$}		
&\multicolumn{2}{l}{$\sigma(gg\to H\to hh_s\to \tau\bar{\tau}b\bar{b})$}		&\multicolumn{2}{l}{$ 2.356 ~\rm fb$} \\

\multicolumn{2}{l}{$\sigma(gg\to H\to h_sh_s\to b\bar{b}b\bar{b})$}			&\multicolumn{2}{l|}{$0.023 ~\rm fb$}		
&\multicolumn{2}{l}{$\sigma(gg\to H\to h_sh_s\to b\bar{b}b\bar{b})$}		&\multicolumn{2}{l}{$1.078 ~\rm fb$} \\

\multicolumn{2}{l}{$\sigma(gg\to A_H\to Zh \to Zb\bar{b}) $}		&\multicolumn{2}{l|}{$7.362 ~\rm fb$}		
&\multicolumn{2}{l}{$\sigma(gg\to A_H\to Zh \to Zb\bar{b}) $}		&\multicolumn{2}{l}{$6.336 ~\rm fb$} \\ 

\multicolumn{2}{l}{$\sigma(gg\to A_H\to Zh_s \to llb\bar{b})$}		&\multicolumn{2}{l|}{$7.454 ~\rm fb$}		
&\multicolumn{2}{l}{$\sigma(gg\to A_H\to Zh_s \to llb\bar{b})$}		&\multicolumn{2}{l}{$8.549 ~\rm fb$} \\ 
\hline
\multicolumn{2}{l}{$C_{h_s gg}, ~C_{h_s VV}, ~C_{h_s \gamma\gamma}, ~C_{h_s t \bar{t}}, ~C_{h_s b\bar{b}}$} 	& \multicolumn{2}{l|}{0.388, ~~0.461, ~~0.386, ~~0.382, ~~0.643}
&\multicolumn{2}{l}{$C_{h_s gg}, ~C_{h_s VV}, ~C_{h_s \gamma\gamma}, ~C_{h_s t \bar{t}}, ~C_{h_s b\bar{b}}$} 	& \multicolumn{2}{l}{0.314,  ~~0.395, ~~0.375, ~~0.314, ~~0.577} \\

 \multicolumn{2}{l}{$C_{h gg}, ~~C_{h VV}, ~~C_{h \gamma\gamma}, ~~C_{h t \bar{t}}, ~~C_{h b\bar{b}}$} 			& \multicolumn{2}{l|}{0.973, ~~0.887, ~~0.909, ~~0.919, ~~0.810}
& \multicolumn{2}{l}{$C_{h gg}, ~~C_{h VV}, ~~C_{h \gamma\gamma}, ~~C_{h t \bar{t}}, ~~C_{h b\bar{b}}$} 		& \multicolumn{2}{l}{0.955,  ~~0.918, ~~0.936, ~~0.947, ~~0.855} \\

 \multicolumn{2}{l}{$C_{H gg}, ~~C_{H VV}, ~~C_{H \gamma\gamma}, ~~C_{H t \bar{t}}, ~~C_{H b\bar{b}}$} 			& \multicolumn{2}{l|}{0.559, ~~0.009, ~~4.001, ~~0.647, ~~1.534}
& \multicolumn{2}{l}{$C_{H gg}, ~~C_{H VV}, ~~C_{H \gamma\gamma}, ~~C_{H t \bar{t}}, ~~C_{H b\bar{b}}$} 		& \multicolumn{2}{l}{0.582,  ~~0.009, ~~3.843, ~~0.673, ~~1.472} \\

\multicolumn{2}{l}{$C_{A_H gg},  ~~C_{A_H \gamma\gamma}, ~~C_{A_H t \bar{t}}, ~~C_{A_H b\bar{b}}$} 			& \multicolumn{2}{l|}{0.650, ~~1.289, ~~0.538, ~~1.305}
& \multicolumn{2}{l}{$C_{A_H gg}, ~~C_{A_H \gamma\gamma}, ~~C_{A_H t \bar{t}}, ~~C_{A_H b\bar{b}}$} 		& \multicolumn{2}{l}{0.673,  ~~1.152, ~~0.545, ~~1.215} \\
\hline
\multicolumn{2}{l}{$N_{11}, ~N_{12}, ~N_{13}, ~N_{14}, ~N_{15}$}   	&\multicolumn{2}{l|}{~0.988,   	~-0.001,  ~~0.102,  ~~0.045,  ~-0.107}
& \multicolumn{2}{l}{$N_{11}, ~N_{12}, ~N_{13}, ~N_{14}, ~N_{15}$}  	&\multicolumn{2}{l}{~0.994, 	~-0.001, ~~0.073,  ~~0.020, ~-0.077} \\

\multicolumn{2}{l}{$N_{21}, ~N_{22}, ~N_{23}, ~N_{24}, ~N_{25}$}   &\multicolumn{2}{l|}{-0.146, 	~~0.009, 	~~0.365,  ~~0.397, ~-0.829}
& \multicolumn{2}{l}{$N_{21}, ~N_{22}, ~N_{23}, ~N_{24}, ~N_{25}$}   &\multicolumn{2}{l}{-0.097,	~~0.006,	 ~~0.300, ~~0.327,  ~-0.891} \\

\multicolumn{2}{l}{$N_{31}, ~N_{32}, ~N_{33}, ~N_{34}, ~N_{35}$}   &\multicolumn{2}{l|}{-0.036,	~-0.583, ~~0.580, ~-0.567,  ~-0.017}
& \multicolumn{2}{l}{$N_{31}, ~N_{32}, ~N_{33}, ~N_{34}, ~N_{35}$}   &\multicolumn{2}{l}{-0.038,	~-0.337, ~~0.668,  ~-0.662,  ~-0.016} \\

\multicolumn{2}{l}{$N_{41}, ~N_{42}, ~N_{43}, ~N_{44}, ~N_{45}$}   &\multicolumn{2}{l|}{~0.029,	~-0.010, ~-0.596, ~-0.586, ~-0.548}
& \multicolumn{2}{l}{$N_{41}, ~N_{42}, ~N_{43}, ~N_{44}, ~N_{45}$}   &\multicolumn{2}{l}{~0.024,	~-0.008,	~-0.636,  ~-0.629, ~-0.447} \\

\multicolumn{2}{l}{$N_{51}, ~N_{52}, ~N_{53}, ~N_{54}, ~N_{55}$}   &\multicolumn{2}{l|}{~0.026,	~-0.812,  ~-0.405,  ~~0.419, ~~0.009}
& \multicolumn{2}{l}{$N_{51}, ~N_{52}, ~N_{53}, ~N_{54}, ~N_{55}$}   &\multicolumn{2}{l}{~0.011,	~-0.941,  ~-0.231,  ~~0.245, ~~0.004} \\
\hline
\multicolumn{2}{l}{ Annihilations }                         & \multicolumn{2}{l|}{Fractions [\%]} 			& \multicolumn{2}{l}{Annihilations}                                       & \multicolumn{2}{l}{Fractions [\%]}                                      \\
\multicolumn{2}{l}{$\tilde{\chi}_1^0\tilde{\chi}_2^0 \to H^\pm W^\mp / h_s A_H /  t\bar{t} / HZ$} 				& \multicolumn{2}{l|}{20.8/13.5/7.5/7.2}
& \multicolumn{2}{l}{$\tilde{\chi}_1^0\tilde{\chi}_1^0 \to h_s A_H / H^\pm W^\mp / t\bar{t} / HZ / h_sZ $} 		& \multicolumn{2}{l}{31.7/26.1/14.7/9.4/2.0}  \\

\multicolumn{2}{l}{$\tilde{\chi}_1^0 \tilde{\chi}_1^0 \to h_sA_H / H^\pm W^\mp /  t\bar{t}/ HZ $} 			& \multicolumn{2}{l|}{11.3/10.3/9.8/7.2}
& \multicolumn{2}{l}{$\tilde{\chi}_1^0 \tilde{\chi}_2^0 \to H^\pm W^\mp / h_sA_H / HZ / t\bar{t} $} 					& \multicolumn{2}{l}{4.2/3.1/1.5/1.0}   \\

\multicolumn{2}{l}{$\tilde{\chi}_2^0  \tilde{\chi}_2^0 \to h_s A_H / t\bar{t} $} 					& \multicolumn{2}{l|}{2.7/2.5}
&    \\
 \hline

\multicolumn{2}{l}{ Decays }                         & \multicolumn{2}{l|}{Branching ratios [\%]} 		& \multicolumn{2}{l}{Decays}                                       & \multicolumn{2}{l}{Branching ratios [\%]}\\
\multicolumn{2}{l}{$h_s \to b \bar{b} / \tau^+ \tau^- / c \bar{c} / gg / W ^+W^{-\ast} / \gamma \gamma$}      & \multicolumn{2}{l|}{87.2/9.65/1.48/1.39/0.10/0.08}
&\multicolumn{2}{l}{$h_s \to b \bar{b}/ \tau^+ \tau^- / c \bar{c} / gg / W ^+W^{-\ast} / \gamma \gamma$}      & \multicolumn{2}{l}{87.6/9.71/1.23/1.16/0.12/0.07}\\

\multicolumn{2}{l}{$h \to b \bar{b} / W ^+W^{-\ast} / \tau^+ \tau^- / gg / ZZ/ c \bar{c} /\gamma \gamma$}      & \multicolumn{2}{l|}{52.1/29.4/6.06/5.53/3.35/3.17/0.029}
&\multicolumn{2}{l}{$h \to b \bar{b} / W ^+W^{-\ast} / \tau^+ \tau^- / gg / ZZ/ c \bar{c} /\gamma \gamma$}      & \multicolumn{2}{l}{58.3/23.2/6.76/5.62/2.44/3.37/0.028}\\

\multicolumn{2}{l}{$H \to  t\bar{t} / A_H Z/ hh_s/ hh$}      & \multicolumn{2}{l|}{72.3/13.3/9.02/4.99}
&\multicolumn{2}{l}{$H \to t\bar{t} / A_H Z/ hh_s/ hh$}      & \multicolumn{2}{l}{65.4/20.6/9.80/3.50}\\

\multicolumn{2}{l}{$A_H \to t\bar{t}/h_s Z /h Z$}     & \multicolumn{2}{l|}{93.7/5.07/0.84}
&\multicolumn{2}{l}{$A_H \to  t\bar{t}/h_s Z /h Z$}    & \multicolumn{2}{l}{94.4/4.70/0.52}\\

\multicolumn{2}{l}{$A_s \to H^\pm W^\mp / h_sA_H / t\bar{t}/ H Z $}      & \multicolumn{2}{l|}{40.4/25.3/16.4/14.3}
&\multicolumn{2}{l}{$A_s \to  H^\pm W^\mp/ h_s A_H / t\bar{t}/H Z$}      & \multicolumn{2}{l}{35.6/ 31.6/16.2/13.2}\\

\multicolumn{2}{l}{$H^+ \to  t \bar{b} / A_H W^+ / h_s W^+ / h W^+$}      & \multicolumn{2}{l|}{81.8/8.6/8.1/1.4}
&\multicolumn{2}{l}{$H^+ \to  t \bar{b} / A_s W^+ / h_s W^+ / h W^+$}      & \multicolumn{2}{l}{74.8/16.7/7.63/0.89}\\

\multicolumn{2}{l}{$\tilde{\chi}^0_2 \to \tilde{\chi}^0_1 Z^\ast$}      & \multicolumn{2}{l|}{100}
&\multicolumn{2}{l}{$\tilde{\chi}^0_2 \to \tilde{\chi}^0_1 Z^\ast$}      & \multicolumn{2}{l}{100} \\

\multicolumn{2}{l}{$\tilde{\chi}^0_3 \to \tilde{\chi}^0_1 Z / \tilde{\chi}^0_2 Z / \tilde{\chi}^0_1 h_s  / \tilde{\chi}^0_1 h$}      & \multicolumn{2}{l|}{61.0/35.7/1.9/1.3}
&\multicolumn{2}{l}{$\tilde{\chi}^0_3 \to \tilde{\chi}^0_2 Z / \tilde{\chi}^0_1 Z  / \tilde{\chi}^0_1 h / \tilde{\chi}^0_1 h_s $}      & \multicolumn{2}{l}{93.7/4.71/1.14/0.22} \\

\multicolumn{2}{l}{$\tilde{\chi}^0_4 \to  \tilde{\chi}^\pm_1 W^\mp /  \tilde{\chi}^0_2 h / \tilde{\chi}^0_2 h_s / \tilde{\chi}^0_3 Z / \tilde{\chi}^0_1 h$}      & \multicolumn{2}{l|}{32.8/36.3/17.0/9.07/2.7}
&\multicolumn{2}{l}{$\tilde{\chi}^0_4 \to  \tilde{\chi}^0_2 h / \tilde{\chi}^0_2 h_s/  \tilde{\chi}^0_1 h / \tilde{\chi}^0_1 Z$}      & \multicolumn{2}{l}{76.2/17.9/2.86/1.57 } \\

\multicolumn{2}{l}{$\tilde{\chi}^0_5 \to \tilde{\chi}^\pm_2 W^\mp / \tilde{\chi}^0_2 Z/ \tilde{\chi}^0_3 h_s /  \tilde{\chi}^0_2 Z $}      & \multicolumn{2}{l|}{41.2/40.7/14.8/2.3}
&\multicolumn{2}{l}{$\tilde{\chi}^0_5 \to \tilde{\chi}^\pm_2 W^\mp / \tilde{\chi}^0_2 Z / \tilde{\chi}^0_3 h_s / \tilde{\chi}^0_4 Z$}      & \multicolumn{2}{l}{47.5/22.6/14.8/6.30} \\

\multicolumn{2}{l}{$\tilde{\chi}^+_1 \to \tilde{\chi}^0_1 W^{+\ast}$}      & \multicolumn{2}{l|}{100}
&\multicolumn{2}{l}{$\tilde{\chi}^+_1 \to \tilde{\chi}^0_1 W^{+\ast}$}      & \multicolumn{2}{l}{100}  \\

\multicolumn{2}{l}{$\tilde{\chi}^+_2 \to \tilde{\chi}^0_2 W^+ / \tilde{\chi}^0_3 W^+ /\tilde{\chi}^+_1 Z /  \tilde{\chi}^+_1 h_s$}      & \multicolumn{2}{l|}{39.6/23.6/19.6/14.7}
&\multicolumn{2}{l}{$\tilde{\chi}^+_2 \to  \tilde{\chi}^0_3 W^+ / \tilde{\chi}^+_1 Z / \tilde{\chi}^0_2 W^+ /  \tilde{\chi}^+_1 h_s$}      & \multicolumn{2}{l}{25.7/23.6/21.2/14.2}\\

\hline
\multicolumn{2}{l}{$S, T, U$}		&\multicolumn{2}{l|}{0.046,  -0.014,  -0.000}
&\multicolumn{2}{l}{$S, T, U$}		&\multicolumn{2}{l}{0.040,   -0.013,  -0.001}\\
\hline

\multicolumn{2}{l}{$R$ value }  & \multicolumn{2}{l|}{$0.24$}
& \multicolumn{2}{l}{$R$ value }  & \multicolumn{2}{l}{$0.23$} \\

\multicolumn{2}{l}{Signal Region}  & \multicolumn{2}{l|}{G05 in CMS-SUS-16-039~\cite{ATLAS:2019lng}}
& \multicolumn{2}{l}{Signal Region}  & \multicolumn{2}{l}{G05 in CMS-SUS-16-039~\cite{ATLAS:2019lng}} \\
\hline \hline

\end{tabular}}
\end{table}

\begin{table}[t]
%\begin{sidewaystable}
\centering
\caption{\label{BP3BP4} Details of the benchmark points P3 from {\bf Scenario I} and P4 from {\bf Scenario II} that are denoted as diamonds colored green and red, respectively, in all presented figures. {  { P3 successfully explains the $\mu_{\gamma\gamma}$ and $\mu_{b\bar{b}}$ excesses at the $2\,\sigma$ level, while slightly exceeding the $2\,\sigma$ preferred region of $\sigma_{\gamma\gamma b\bar{b}}$.
P4 yields a $2\,\sigma$ interpretation of the combined $\sigma_{\gamma\gamma b\bar{b}}$ and $\mu_{\gamma\gamma}$ anomalies, with $\mu_{b\bar{b}}$ falling within the $2\,\sigma$ interval as well.}}  In the listed annihilation final states, $d_i$ and $u_i$ denote the $i$th generation of down-type and up-type quarks, respectively.}
\vspace{0.3cm}

\centering
\resizebox{1\textwidth}{!}
 {
\begin{tabular}{lrlr|lrlr}
\hline \hline
\multicolumn{4}{c|}{\bf Benchmark Point P3}                                                                                                				& \multicolumn{4}{c}{\bf Benchmark Point P4}
 \\ \hline
$\mu_{tot}$                		& 600.8~GeV 	&$\lambda$             		& 0.622				&$\mu_{tot}$                & 711.8~GeV 	&$\lambda$             		& 0.688 \\
$\mu$					& 277.1~GeV	&$\kappa$              		& 0.213				&$\mu$				& 581.5~GeV	&$\kappa$              		& -0.094\\
$\mu^\prime$			& -703.3~GeV	&$\delta$         			& -0.001				&$\mu^\prime$		& 750.4~GeV	&$\delta$          			& 0.033\\
$M_1$               			& -378.0~GeV 	&$\tan{\beta}$                	& 1.442				&$M_1$               		& -656.4~GeV 	&$\tan{\beta}$                	& 1.467\\
$M_2$             		 	& 532.8~GeV	&$A_{\lambda}$	 		&1765~GeV			&$M_2$             		& 855.4~GeV	&$A_{\lambda}$ 		       & 763.8~GeV	\\
$m_A$					& 641.1~GeV	&$A_{\kappa}$			&1702~GeV			&$m_A$				&616.0~GeV		&$A_{\kappa}$			& -1955~GeV  \\
$m_B$					& 110.0~GeV	&$A_t$					&1441~GeV			&$m_B$				&97.57~GeV		&$A_t$					&1082~GeV	\\
$m_N$					& -481.0~GeV	&						&					&$m_N$				&714.9~GeV	&						&   \\
\hline
$m_{\tilde{\chi}_1^0}$   	 & -381.7~GeV	&$m_{h_s}$                	& 94.86~GeV		&$m_{\tilde{\chi}_1^0}$   		& -660.9~GeV	&$m_{h_s}$                		& 95.37~GeV\\
$m_{\tilde{\chi}_2^0}$   	 & -416.1~GeV	&$m_{h}$              	& 126.4~GeV		&$m_{\tilde{\chi}_2^0}$   		& 679.0~GeV	&$m_{h}$              		& 127.3~GeV\\
$m_{\tilde{\chi}_3^0}$   	 & 502.3~GeV	&$m_{H}$                  	& 661.1~GeV		&$m_{\tilde{\chi}_3^0}$   		& -728.8~GeV	&$m_{H}$                  		& 629.0~GeV\\
$m_{\tilde{\chi}_4^0}$   	 & 661.4~GeV	&$m_{A_H}$                & 514.1~GeV			&$m_{\tilde{\chi}_4^0}$   		& 730.3~GeV	&$m_{A_H}$                 	& 615.6~GeV\\
$m_{\tilde{\chi}_5^0}$   	 & -662~GeV		&$m_{A_s}$               & 876.3~GeV			&$m_{\tilde{\chi}_5^0}$   		& 913.3~GeV	&$m_{A_s}$                	& 809.9~GeV\\
$m_{\tilde{\chi}_1^\pm}$	 & 501.7~GeV	&$m_{H^\pm}$		& 650.0~GeV		&$m_{\tilde{\chi}_1^\pm}$		& 688.9~GeV	&$m_{H^\pm}$			& 617.9~GeV\\
$m_{\tilde{\chi}_2^\pm}$	 & 661.2~GeV	& 					& 					&$m_{\tilde{\chi}_2^\pm}$	 	& 913.1~GeV	&						& \\
\hline
\multicolumn{2}{l}{$\Omega h^2 $}			&\multicolumn{2}{l|}{0.104}
&\multicolumn{2}{l}{$\Omega h^2 $}		&\multicolumn{2}{l}{0.114}\\
\multicolumn{2}{l}{$\sigma^{SI}_p$,  ~$\sigma^{SD}_n$}		&\multicolumn{2}{l|}{$1.1\times 10^{-50}{\rm ~cm^2}, ~1.6\times 10^{-42}{\rm ~~cm^2}$}
&\multicolumn{2}{l}{$\sigma^{SI}_p$, ~$\sigma^{SD}_n$}		&\multicolumn{2}{l}{$7.4\times 10^{-48}{\rm ~cm^2},  ~1.7\times 10^{-42}{\rm ~~cm^2}$}\\
\hline
\multicolumn{2}{l}{$\mu_{\gamma\gamma}$}		&\multicolumn{2}{l|}{0.088}		
&\multicolumn{2}{l}{$\mu_{\gamma\gamma}$}	&\multicolumn{2}{l}{0.195} \\ 

\multicolumn{2}{l}{$ \mu_{b\bar{b}}$}		&\multicolumn{2}{l|}{0.225}		
&\multicolumn{2}{l}{$\mu_{b\bar{b}}$}		&\multicolumn{2}{l}{0.014} \\ 

\multicolumn{2}{l}{$\sigma_{\gamma\gamma b\bar{b}}$}		&\multicolumn{2}{l|}{$0.074 ~\rm fb$}		
&\multicolumn{2}{l}{$\sigma_{\gamma\gamma b\bar{b}}$}		&\multicolumn{2}{l}{$0.045 ~\rm fb$} \\ 

\multicolumn{2}{l}{$\sigma(gg\to H\to hh_s\to \tau\bar{\tau}b\bar{b}) $}		&\multicolumn{2}{l|}{$1.632 ~\rm fb$}		
&\multicolumn{2}{l}{$\sigma(gg\to H\to hh_s\to \tau\bar{\tau}b\bar{b})$}		&\multicolumn{2}{l}{$1.262 ~\rm fb$} \\

\multicolumn{2}{l}{$\sigma(gg\to H\to h_sh_s\to b\bar{b}b\bar{b})$}			&\multicolumn{2}{l|}{$0.071 ~\rm fb$}		
&\multicolumn{2}{l}{$\sigma(gg\to H\to h_sh_s\to b\bar{b}b\bar{b})$}		&\multicolumn{2}{l}{$36.75 ~\rm fb$} \\

\multicolumn{2}{l}{$\sigma(gg\to A_H\to Zh \to Zb\bar{b}) $}		&\multicolumn{2}{l|}{$5.086 ~\rm fb$}		
&\multicolumn{2}{l}{$\sigma(gg\to A_H\to Zh \to Zb\bar{b}) $}		&\multicolumn{2}{l}{$0.579 ~\rm fb$} \\ 

\multicolumn{2}{l}{$\sigma(gg\to A_H\to Zh_s \to llb\bar{b})$}		&\multicolumn{2}{l|}{$5.414 ~\rm fb$}		
&\multicolumn{2}{l}{$\sigma(gg\to A_H\to Zh_s \to llb\bar{b})$}		&\multicolumn{2}{l}{$0.966 ~\rm fb$} \\ 
\hline
\multicolumn{2}{l}{$C_{h_s gg}, ~C_{h_s VV}, ~C_{h_s \gamma\gamma}, ~C_{h_s t \bar{t}}, ~C_{h_s b\bar{b}}$} 	& \multicolumn{2}{l|}{0.391, ~~0.456, ~~0431, ~~0.386, ~~0.603}
&\multicolumn{2}{l}{$C_{h_s gg}, ~C_{h_s VV}, ~C_{h_s \gamma\gamma}, ~C_{h_s t \bar{t}}, ~C_{h_s b\bar{b}}$} 	& \multicolumn{2}{l}{0.280,  ~~0.145, ~~0.172, ~~0.254, ~~0.089} \\

 \multicolumn{2}{l}{$C_{h gg}, ~~C_{h VV}, ~~C_{h \gamma\gamma}, ~~C_{h t \bar{t}}, ~~C_{h b\bar{b}}$} 			& \multicolumn{2}{l|}{0.924, ~~0.889, ~~0.914, ~~0.918, ~~0.830}
& \multicolumn{2}{l}{$C_{h gg}, ~~C_{h VV}, ~~C_{h \gamma\gamma}, ~~C_{h t \bar{t}}, ~~C_{h b\bar{b}}$} 		& \multicolumn{2}{l}{0.967,  ~~0.989, ~~1.006, ~~0.966, ~~1.039} \\

 \multicolumn{2}{l}{$C_{H gg}, ~~C_{H VV}, ~~C_{H \gamma\gamma}, ~~C_{H t \bar{t}}, ~~C_{H b\bar{b}}$} 			& \multicolumn{2}{l|}{0.604, ~~0.010, ~~4.129, ~~0.699, ~~1.423}
& \multicolumn{2}{l}{$C_{H gg}, ~~C_{H VV}, ~~C_{H \gamma\gamma}, ~~C_{H t \bar{t}}, ~~C_{H b\bar{b}}$} 		& \multicolumn{2}{l}{0.593,  ~~0.011, ~~4.530, ~~0.683, ~~1.437} \\

\multicolumn{2}{l}{$C_{A_H gg},  ~~C_{A_H \gamma\gamma}, ~~C_{A_H t \bar{t}}, ~~C_{A_H b\bar{b}}$} 			& \multicolumn{2}{l|}{0.683, ~~1.380, ~~0.568, ~~1.181}
& \multicolumn{2}{l}{$C_{A_H gg}, ~~C_{A_H \gamma\gamma}, ~~C_{A_H t \bar{t}}, ~~C_{A_H b\bar{b}}$} 		& \multicolumn{2}{l}{0.745,  ~~4.688, ~~0.681, ~~1.467} \\
\hline
\multicolumn{2}{l}{$N_{11}, ~N_{12}, ~N_{13}, ~N_{14}, ~N_{15}$}   	&\multicolumn{2}{l|}{~0.988,   	~-0.002,  ~~0.034,  ~~0.032,  ~-0.116}
& \multicolumn{2}{l}{$N_{11}, ~N_{12}, ~N_{13}, ~N_{14}, ~N_{15}$}  	&\multicolumn{2}{l}{-0.992, 	~~0.000, ~-0.108,  ~-0.064, ~-0.001} \\

\multicolumn{2}{l}{$N_{21}, ~N_{22}, ~N_{23}, ~N_{24}, ~N_{25}$}   &\multicolumn{2}{l|}{-0.144, 	~~0.008, 	~~0.318,  ~~0.351, ~-0.869}
& \multicolumn{2}{l}{$N_{21}, ~N_{22}, ~N_{23}, ~N_{24}, ~N_{25}$}   &\multicolumn{2}{l}{-0.113,	~-0.027,	 ~-0.317, ~-0.613,  ~~0.452} \\

\multicolumn{2}{l}{$N_{31}, ~N_{32}, ~N_{33}, ~N_{34}, ~N_{35}$}   &\multicolumn{2}{l|}{-0.028,	~-0.815, ~~0.416, ~-0.401,  ~-0.012}
& \multicolumn{2}{l}{$N_{31}, ~N_{32}, ~N_{33}, ~N_{34}, ~N_{35}$}   &\multicolumn{2}{l}{-0.122,	~~0.010,  ~~0.698,  ~~0.702,  ~~0.008} \\

\multicolumn{2}{l}{$N_{41}, ~N_{42}, ~N_{43}, ~N_{44}, ~N_{45}$}   &\multicolumn{2}{l|}{~0.034,	~-0.578, ~-0.572, ~~0.580, ~~0.013}
& \multicolumn{2}{l}{$N_{41}, ~N_{42}, ~N_{43}, ~N_{44}, ~N_{45}$}   &\multicolumn{2}{l}{-0.012,	~-0.200,	~~0.340,  ~-0.237, ~-0.888} \\

\multicolumn{2}{l}{$N_{51}, ~N_{52}, ~N_{53}, ~N_{54}, ~N_{55}$}   &\multicolumn{2}{l|}{~0.023,	~-0.010,  ~-0.623,  ~-0.615, ~-0.481}
& \multicolumn{2}{l}{$N_{51}, ~N_{52}, ~N_{53}, ~N_{54}, ~N_{55}$}   &\multicolumn{2}{l}{~0.010,	~-0.927,  ~-0.260,  ~~0.268, ~~0.038} \\
\hline
\multicolumn{2}{l}{ Annihilations }                         & \multicolumn{2}{l|}{Fractions [\%]} 			& \multicolumn{2}{l}{Annihilations}                                       & \multicolumn{2}{l}{Fractions [\%]}                                      \\
\multicolumn{2}{l}{$\tilde{\chi}_2^0\tilde{\chi}_2^0 \to  h_s A_H / H^\pm W^\mp / t\bar{t} / HZ$} 				& \multicolumn{2}{l|}{26.3/25.4/12.1/9.7}
& \multicolumn{2}{l}{$\tilde{\chi}_2^0\tilde{\chi}_1^- \to d_i \bar{u}_i /  h_s W^- / h W^- / A_s W^- /A_H W^- $} 	& \multicolumn{2}{l}{11/4.4/2.9/2.9/1.6}  \\

\multicolumn{2}{l}{$\tilde{\chi}_1^0 \tilde{\chi}_2^0 \to h_sA_H / H^\pm W^\mp /  t\bar{t}/ HZ $} 			& \multicolumn{2}{l|}{5.2/4.3/3.8/1.4}
& \multicolumn{2}{l}{$\tilde{\chi}_2^0\tilde{\chi}_2^0 \to h_sA_s /  HZ/ t\bar{t}/ W^+ W^- $} 		& \multicolumn{2}{l}{5.7/1.7/1.6/1.5}   \\

\multicolumn{2}{l}{$\tilde{\chi}_1^0  \tilde{\chi}_1^0 \to  t\bar{t} / h_s A_H$} 					& \multicolumn{2}{l|}{2.9/2.7}
& \multicolumn{2}{l}{$\tilde{\chi}_1^\pm \tilde{\chi}_1^\mp \to h_s A_s /  W^+ W^-/ t\bar{t}$} 					& \multicolumn{2}{l}{4.8/1.4/1.3}   \\
 \hline

\multicolumn{2}{l}{ Decays }                         & \multicolumn{2}{l|}{Branching ratios [\%]} 		& \multicolumn{2}{l}{Decays}                                       & \multicolumn{2}{l}{Branching ratios [\%]}\\
\multicolumn{2}{l}{$h_s \to b \bar{b} / \tau^+ \tau^- / c \bar{c} / gg / W ^+W^{-\ast} / \gamma \gamma$}      & \multicolumn{2}{l|}{86.8/9.61/1.68/1.59/0.11/0.08}
&\multicolumn{2}{l}{$h_s \to b \bar{b}/ gg / c \bar{c} / \tau^+ \tau^-  /  W ^+W^{-\ast} / \gamma \gamma$}      & \multicolumn{2}{l}{51.4/22.5/19.7/5.69/0.37/0.36}\\

\multicolumn{2}{l}{$h \to b \bar{b} / W ^+W^{-\ast} / \tau^+ \tau^- / gg / c \bar{c} / ZZ/\gamma \gamma$}      & \multicolumn{2}{l|}{54.3/27.4/6.31/5.41/3.13/3.08/0.28}
&\multicolumn{2}{l}{$h \to b \bar{b} / W ^+W^{-\ast} / \tau^+ \tau^- / gg / ZZ/ c \bar{c} /\gamma \gamma$}      & \multicolumn{2}{l}{57.9/25.7/6.74/4.10/2.91/2.36/0.31}\\

\multicolumn{2}{l}{$H \to  t\bar{t} / A_H Z/ hh_s/ hh$}      & \multicolumn{2}{l|}{76.5/12.8/6.6/3.8}
&\multicolumn{2}{l}{$H \to t\bar{t} / h_sh_s/ hh_s/hh$}      & \multicolumn{2}{l}{67.8/25.1/6.6/0.20}\\

\multicolumn{2}{l}{$A_H \to t\bar{t}/h_s Z /h Z$}     & \multicolumn{2}{l|}{95.8/3.4/0.5}
&\multicolumn{2}{l}{$A_H \to  t\bar{t}/h_s Z /h Z$}    & \multicolumn{2}{l}{87.5/11.7/0.5}\\

\multicolumn{2}{l}{$A_s \to H^\pm W^\mp / h_sA_H /t\bar{t}/ h_s Z $}      & \multicolumn{2}{l|}{33.2/29.6/19.4/2.2}
&\multicolumn{2}{l}{$A_s \to h_s A_H / H^\pm W^\mp/ t\bar{t}/h A_H$}      & \multicolumn{2}{l}{39.6/20.9/18.7/8.6}\\

\multicolumn{2}{l}{$H^+ \to  t \bar{b} / A_H W^+ / h_s W^+ / h W^+$}      & \multicolumn{2}{l|}{82.5/11.0/5.6/0.9}
&\multicolumn{2}{l}{$H^+ \to   t \bar{b} / h_s W^+ / h W^+$}      & \multicolumn{2}{l}{85.9/13.5/0.59}\\

\multicolumn{2}{l}{$\tilde{\chi}^0_2 \to \tilde{\chi}^0_1 Z^\ast$}      & \multicolumn{2}{l|}{100}
&\multicolumn{2}{l}{$\tilde{\chi}^0_2 \to \tilde{\chi}^0_1 Z^\ast$}      & \multicolumn{2}{l}{100} \\

\multicolumn{2}{l}{$\tilde{\chi}^0_3 \to \tilde{\chi}^0_1 Z /  \tilde{\chi}^0_1 h_s$}      & \multicolumn{2}{l|}{74.1/3.5}
&\multicolumn{2}{l}{$\tilde{\chi}^0_3 \to \tilde{\chi}^0_2 Z / \tilde{\chi}^\pm_1 W^{\mp} /  \tilde{\chi}^0_1 Z$}      & \multicolumn{2}{l}{49.8/48.6/3.13} \\

\multicolumn{2}{l}{$\tilde{\chi}^0_4 \to  \tilde{\chi}^0_2 Z/ \tilde{\chi}^\pm_1 W^\mp / \tilde{\chi}^0_3 h /  \tilde{\chi}^0_3 h_s$}      & \multicolumn{2}{l|}{48.6/36.9/6.4/5.1}
&\multicolumn{2}{l}{$\tilde{\chi}^0_4 \to \tilde{\chi}^\pm_1 W^\mp /\tilde{\chi}^0_1 h_s / \tilde{\chi}^0_2 Z$}      & \multicolumn{2}{l}{69.7/15.6/5.80 } \\

\multicolumn{2}{l}{$\tilde{\chi}^0_5 \to \tilde{\chi}^\pm_1 W^\mp /  \tilde{\chi}^0_2 h / \tilde{\chi}^0_3 Z / \tilde{\chi}^0_2 h_s$}      & \multicolumn{2}{l|}{39.5/32.9/16.5/8.2}
&\multicolumn{2}{l}{$\tilde{\chi}^0_5 \to \tilde{\chi}^\pm_2 W^\mp / \tilde{\chi}^0_3 Z / \tilde{\chi}^0_2 h / \tilde{\chi}^0_1 h_s$}      & \multicolumn{2}{l}{55.7/16.7/15.9/6.61} \\

\multicolumn{2}{l}{$\tilde{\chi}^+_1 \to \tilde{\chi}^0_1 W^{+\ast} / \tilde{\chi}^0_2 W^{+\ast}$}      & \multicolumn{2}{l|}{60.6/ 39.4}
&\multicolumn{2}{l}{$\tilde{\chi}^+_1 \to \tilde{\chi}^0_1 W^{+\ast}/ \tilde{\chi}^0_2 W^{+\ast}$}      & \multicolumn{2}{l}{75.2/24.7}  \\

\multicolumn{2}{l}{$\tilde{\chi}^+_2 \to \tilde{\chi}^0_2 W^+ / \tilde{\chi}^0_3 W^+ /\tilde{\chi}^+_1 Z /  \tilde{\chi}^+_1 h / \tilde{\chi}^+_1 h_s$}      & \multicolumn{2}{l|}{47.7/20.7/17.1/6.4/5.1}
&\multicolumn{2}{l}{$\tilde{\chi}^+_2 \to \tilde{\chi}^+_1 Z / \tilde{\chi}^0_2 W^+ / \tilde{\chi}^+_1 h /    \tilde{\chi}^0_3 W^+ / \tilde{\chi}^0_4 W^+$}      & \multicolumn{2}{l}{27.4/22.4/19.4/17.3/6.43}\\
\hline
\multicolumn{2}{l}{$S, T, U$}		&\multicolumn{2}{l|}{0.033,  -0.010,  -0.000}
&\multicolumn{2}{l}{$S, T, U$}		&\multicolumn{2}{l}{0.279,   -0.051,   0.008}\\
\hline

\multicolumn{2}{l}{$R$ value }  & \multicolumn{2}{l|}{$0.21$}
& \multicolumn{2}{l}{$R$ value }  & \multicolumn{2}{l}{$0.02$} \\

\multicolumn{2}{l}{Signal Region}  & \multicolumn{2}{l|}{G05 in CMS-SUS-16-039~\cite{ATLAS:2019lng}}
& \multicolumn{2}{l}{Signal Region}  & \multicolumn{2}{l}{{SR2-stop-3high-pt-1} in CMS-SUS-16-048~\cite{CMS:2017moi}} \\
\hline \hline

\end{tabular}}
\end{table}

{   {In addition, we further investigate the constraints from electroweak precision observables on the interpretation of the observed anomalies. For each viable parameter point, the oblique parameters S, T and U are evaluated using the \textsf{SPheno} package and confronted with the results of the global electroweak fit~\cite{ParticleDataGroup:2024cfk}:
\begin{equation}
S = 0.021 \pm 0.096, \qquad
T = 0.040 \pm 0.120, \qquad
U = 0.008 \pm 0.092.
\end{equation}
with a strong positive correlation between $S$ and $T$, $\rho_{ST} = 0.91$,
and negative correlations between $S$ and $U$,  $\rho_{SU} = -0.62$,
as well as between $T$ and $U$, $\rho_{TU} = -0.83$.
We calculate the $\chi^2$ value assuming three degrees of freedom and require agreement with the fit result within a confidence level of $2\,\sigma$. Under this criterion, the allowed ranges of the oblique parameters relevant for the interpretation of the anomalies are found to be $0.013 \lesssim S  \lesssim 0.049$, $-0.017 \lesssim T  \lesssim 0.0002$, $U  \simeq 0.0$.
We find that electroweak precision tests exclude only a small fraction of the parameter space in {\bf Scenario I}, without significantly altering the overall phenomenological picture. In this scenario, the mass splitting between the charged Higgs bosons and the neutral CP-odd Higgs boson, $\Delta m (H^\pm , A_H)$, can reach values as large as about 170 GeV.  This feature can be understood by noting that {\bf Scenario I} allows for relatively large mixing induced by the enhancement of $A_\lambda - m_N$ (see the discussion of the lower-left panel of Fig.~\ref{Fig3}). We present three representative benchmark points, P1, P2, and P3 in Table.~\ref{BP1BP2}, which satisfy all electroweak precision constraints in {\bf Scenario I}.
In contrast, the oblique parameters impose much more stringent constraints on \textbf{Scenario II}, excluding more than half of the sampled parameter points. As an illustrative example of an excluded configuration, we provide the detailed properties of benchmark point P4 in Table.~\ref{BP3BP4}, which fails to satisfy the electroweak precision bounds.}}

\section{Conclusion} \label{conclusion}
The $\mu_{\gamma\gamma}$ excess reported by CMS and ATLAS and the $\mu_{b\bar{b}}$ excess observed at LEP  hint in a compatible way at the existence of a light Higgs boson near 95 GeV. In parallel, an excess in the resonant production of SM-like plus BSM Higgs bosons in the diphoton plus $b\bar{b}$ channel by CMS suggests an extra heavy Higgs bosons near 650 GeV. 
This research rigorously investigated the capacity of the GNMSSM to simultaneously explain the 95 GeV and 650 GeV excesses while providing a proper DM candidate with correct DM relic abundance. 
We adopt a framework in which the light singlet-like $CP$-even Higgs boson $h_s$ accounts for the the 95 GeV anomalies, whereas the heavy doublet-like $CP$-even Higgs boson $H$ decays into the SM-like Higgs $h$ and $h_s$ to account for the $\sigma_{\gamma\gamma b\bar{b}}$ excess. 
This configuration, characterized by small $\tan\beta$ and large $\lambda$, is theoretically well-motivated as it leverages the tree-level singlet-doublet mixing to elevate the mass of the $125 \text{ GeV}$ Higgs boson without relying on excessively large radiative corrections.
To streamline the parameterization, a small auxiliary factor $\delta$ is introduced, which substantially improves sampling efficiency without affecting the underlying physics.
By performing a comprehensive 12-dimensional parameter scan using state-of-the-art sampling and numerical tools, constrained by a battery of experimental limits including $125 \text{ GeV}$ Higgs data, $B$-physics, vacuum stability, and the most stringent LZ detection results, we successfully demonstrated the model's viability.
The key findings and insights derived from this analysis are summarized below:
\begin{enumerate}
    \item The best fit in {\bf Scenario I} provides a simultaneous interpretation of the $\sigma_{\gamma\gamma b\bar{b}}$ excess near 650 GeV and $\mu_{\gamma\gamma}$ and $\mu_{b\bar{b}}$ excesses near 95 GeV at the $2\,\sigma$ level. Larger values of $\sigma_{\gamma\gamma b\bar{b}} \gtrsim 0.1 ~\rm fb$  are limited by the CMS search in the $\tau\bar{\tau}$ plus $b\bar{b}$ channel.
    \item {\bf Scenario II} successfully accommodates the combined $\sigma_{\gamma\gamma b\bar{b}}$ and $\mu_{\gamma\gamma}$ excesses at the $2\,\sigma$ level, with $\mu_{\gamma\gamma}$ achieving its central value while $\mu_{b\bar{b}}$ remains within the $2\,\sigma$ interval.  However, electroweak precision observables impose significantly stronger constraints on this scenario.
    \item The consistent interpretation of the scalar anomalies robustly favors the Bino-dominated $\tilde{\chi}^0_1$ as the sole DM candidate. The measured relic density is achieved through distinct mechanisms in each scenario: {\bf Scenario I} primarily relies on $A_s$ funnel annihilation or coannihilation with $\tilde{S}$-like $\tilde{\chi}^0_2$s, and {\bf Scenario II} favors coannihilation with $\tilde{H}$-like $\tilde{\chi}^0_2$s or $\tilde{\chi}^\pm_1$s.
    \item  Constraints from LHC searches for electroweakinos have a negligible impact on the viable parameter space. Full Monte Carlo simulations for benchmark points consistently yield low $R$-values ($< 0.30$). This is attributed to the relatively heavy electroweakino masses, which reduce production rates, and the compressed mass spectra, which lead to detection-challenging soft final states.
    \item  The necessary non-zero values for the bilinear parameters $\mathbf{\mu}$ and $\mathbf{\mu^\prime}$ serve as a direct signature that distinguishes the GNMSSM from the minimal $\mathbb{Z}_3$-NMSSM and the MSSM.
\end{enumerate}

Beyond these results, we provide prospects for several promising search channels targeting a heavy Higgs boson near 650 GeV, presenting posterior probability density functions for key cross-sections and normalized couplings. These projections offer concrete guidance for testing or excluding the GNMSSM parameter space in future analyses. In addition, the benchmark points identified in this work enable reliable extrapolations to the expected sensitivities of the High-Luminosity LHC.
Overall, our results highlight the GNMSSM as a well-motivated extension of the Standard Model that can consistently account for both the observed scalar excesses and a viable DM candidate. This motivates continued experimental efforts in forthcoming collider programs and DM detection experiments.

\section*{Acknowledgements}
We thank Prof. Junjie Cao for helpful discussions. This work is supported by the National Natural Science Foundation of China (NNSFC) under Grant No.12447140 and the Natural Science Foundation of Henan Province, China (Grant No. 252300421771, 252300421988)

\clearpage
\appendix 
\section{Appendix}

% 表格从 1 开始编号
\setcounter{table}{0}
\renewcommand{\thetable}{A\arabic{table}}

\begin{table}[H]
\caption{\label{LHCanalyses} Experimental analyses of the electroweakino production processes considered in this study categorizing by the topologies of the SUSY signals.} 
\label{tab:LHC}
	\vspace{0.2cm}
	\resizebox{0.97\textwidth}{!}{
		\begin{tabular}{llll}
			\hline\hline
			\texttt{Scenario} & \texttt{Final State} &\multicolumn{1}{c}{\texttt{Name}}\\\hline
			\multirow{6}{*}{$\tilde{\chi}_{2}^0\tilde{\chi}_1^{\pm}\rightarrow WZ\tilde{\chi}_1^0\tilde{\chi}_1^0$}&\multirow{6}{*}{$n\ell (n\geq2) + nj(n\geq0) + \text{E}_\text{T}^{\text{miss}}$}&\texttt{CMS-SUS-20-001($137fb^{-1}$)}~\cite{CMS:2020bfa}\\&&\texttt{ATLAS-2106-01676($139fb^{-1}$)}~\cite{ATLAS:2021moa}\\&&\texttt{CMS-SUS-17-004($35.9fb^{-1}$)}~\cite{CMS:2018szt}\\&&\texttt{CMS-SUS-16-039($35.9fb^{-1}$)}~\cite{CMS:2017moi}\\&&\texttt{ATLAS-1803-02762($36.1fb^{-1}$)}~\cite{ATLAS:2018ojr}\\&&\texttt{ATLAS-1806-02293($36.1fb^{-1}$)}~\cite{ATLAS:2018eui}\\\\
			\multirow{6}{*}{$\tilde{\chi}_{2}^0\tilde{\chi}_1^{\pm}\rightarrow Wh\tilde{\chi}_1^0\tilde{\chi}_1^0$}&\multirow{6}{*}{$n\ell(n\geq1) + nb(n\geq0) + nj(n\geq0) + \text{E}_\text{T}^{\text{miss}}$}&\texttt{ATLAS-1909-09226($139fb^{-1}$)}~\cite{ATLAS:2020pgy}\\&&\texttt{CMS-SUS-17-004($35.9fb^{-1}$)}~\cite{CMS:2018szt}\\&&\texttt{CMS-SUS-16-039($35.9fb^{-1}$)}~\cite{CMS:2017moi}\\
			&&\texttt{ATLAS-1812-09432($36.1fb^{-1}$)}\cite{ATLAS:2018qmw}\\&&\texttt{CMS-SUS-16-034($35.9fb^{-1}$)}\cite{CMS:2017kxn}\\&&\texttt{CMS-SUS-16-045($35.9fb^{-1}$)}~\cite{CMS:2017bki}\\\\
			\multirow{2}{*}{$\tilde{\chi}_1^{\mp}\tilde{\chi}_1^{\pm}\rightarrow WW\tilde{\chi}_1^0 \tilde{\chi}_1^0$}&\multirow{2}{*}{$2\ell + \text{E}_\text{T}^{\text{miss}}$}&\texttt{ATLAS-1908-08215($139fb^{-1}$)}~\cite{ATLAS:2019lff}\\&&\texttt{CMS-SUS-17-010($35.9fb^{-1}$)}~\cite{CMS:2018xqw}\\\\
			\multirow{1}{*}{$\tilde{\chi}_2^{0}\tilde{\chi}_1^{\pm}\rightarrow ZW\tilde{\chi}_1^0\tilde{\chi}_1^0$}&\multirow{2}{*}{$2j(\text{large}) + \text{E}_\text{T}^{\text{miss}}$}&\multirow{2}{*}{\texttt{ATLAS-2108-07586($139fb^{-1}$)}~\cite{ATLAS:2021yqv}}\\
			{$\tilde{\chi}_1^{\pm}\tilde{\chi}_1^{\mp}\rightarrow WW\tilde{\chi}_1^0\tilde{\chi}_1^0$}&&\\\\
			\multirow{1}{*}{$\tilde{\chi}_2^{0}\tilde{\chi}_1^{\pm}\rightarrow (h/Z)W\tilde{\chi}_1^0\tilde{\chi}_1^0$}&\multirow{2}{*}{$j(\text{large}) + b(\text{large}) + \text{E}_\text{T}^{\text{miss}}$}&\multirow{2}{*}{\texttt{ATLAS-2108-07586($139fb^{-1}$)}~\cite{ATLAS:2021yqv}}\\
			{$\tilde{\chi}_2^{0}\tilde{\chi}_3^{0}\rightarrow (h/Z)Z\tilde{\chi}_1^0\tilde{\chi}_1^0$}&&\\\\
			$\tilde{\chi}_2^{0}\tilde{\chi}_1^{\mp}\rightarrow h/ZW\tilde{\chi}_1^0\tilde{\chi}_1^0,\tilde{\chi}_1^0\rightarrow \gamma/Z\tilde{G}$&\multirow{2}{*}{$2\gamma + n\ell(n\geq0) + nb(n\geq0) + nj(n\geq0) + \text{E}_\text{T}^{\text{miss}}$}&\multirow{2}{*}{\texttt{ATLAS-1802-03158($36.1fb^{-1}$)}~\cite{ATLAS:2018nud}}\\$\tilde{\chi}_1^{\pm}\tilde{\chi}_1^{\mp}\rightarrow WW\tilde{\chi}_1^0\tilde{\chi}_1^0,\tilde{\chi}_1^0\rightarrow \gamma/Z\tilde{G}$&&\\\\
			$\tilde{\chi}_2^{0}\tilde{\chi}_1^{\pm}\rightarrow ZW\tilde{\chi}_1^0\tilde{\chi}_1^0,\tilde{\chi}_1^0\rightarrow h/Z\tilde{G}$&\multirow{4}{*}{$n\ell(n\geq4) + \text{E}_\text{T}^{\text{miss}}$}&\multirow{4}{*}{\texttt{ATLAS-2103-11684($139fb^{-1}$)}~\cite{ATLAS:2021yyr}}\\$\tilde{\chi}_1^{\pm}\tilde{\chi}_1^{\mp}\rightarrow WW\tilde{\chi}_1^0\tilde{\chi}_1^0,\tilde{\chi}_1^0\rightarrow h/Z\tilde{G}$&&\\$\tilde{\chi}_2^{0}\tilde{\chi}_1^{0}\rightarrow Z\tilde{\chi}_1^0\tilde{\chi}_1^0,\tilde{\chi}_1^0\rightarrow h/Z\tilde{G}$&&\\$\tilde{\chi}_1^{\mp}\tilde{\chi}_1^{0}\rightarrow W\tilde{\chi}_1^0\tilde{\chi}_1^0,\tilde{\chi}_1^0\rightarrow h/Z\tilde{G}$&&\\\\
			\multirow{3}{*}{$\tilde{\chi}_{i}^{0,\pm}\tilde{\chi}_{j}^{0,\mp}\rightarrow \tilde{\chi}_1^0\tilde{\chi}_1^0+\chi_{soft}\rightarrow ZZ/H\tilde{G}\tilde{G}$}&\multirow{3}{*}{$n\ell(n\geq2) + nb(n\geq0) + nj(n\geq0) + \text{E}_\text{T}^{\text{miss}}$}&\texttt{CMS-SUS-16-039($35.9fb^{-1}$)}~\cite{CMS:2017moi}\\&&\texttt{CMS-SUS-17-004($35.9fb^{-1}$)}~\cite{CMS:2018szt}\\&&\texttt{CMS-SUS-20-001($137fb^{-1}$)}~\cite{CMS:2020bfa}\\\\
			\multirow{2}{*}{$\tilde{\chi}_{i}^{0,\pm}\tilde{\chi}_{j}^{0,\mp}\rightarrow \tilde{\chi}_1^0\tilde{\chi}_1^0+\chi_{soft}\rightarrow HH\tilde{G}\tilde{G}$}&\multirow{2}{*}{$n\ell(n\geq2) + nb(n\geq0) + nj(n\geq0) + \text{E}_\text{T}^{\text{miss}}$}&\texttt{CMS-SUS-16-039($35.9fb^{-1}$)}~\cite{CMS:2017moi}\\&&\texttt{CMS-SUS-17-004($35.9fb^{-1}$)}~\cite{CMS:2018szt}\\\\
			$\tilde{\chi}_{2}^{0}\tilde{\chi}_{1}^{\pm}\rightarrow W^{*}Z^{*}\tilde{\chi}_1^0\tilde{\chi}_1^0$&$3\ell + \text{E}_\text{T}^{\text{miss}}$&\texttt{ATLAS-2106-01676($139fb^{-1}$)}~\cite{ATLAS:2021moa}\\\\
			\multirow{3}{*}{$\tilde{\chi}_{2}^{0}\tilde{\chi}_{1}^{\pm}\rightarrow Z^{*}W^{*}\tilde{\chi}_1^0\tilde{\chi}_1^0$}&\multirow{2}{*}{$2\ell + nj(n\geq0) + \text{E}_\text{T}^{\text{miss}}$}&\texttt{ATLAS-1911-12606($139fb^{-1}$)}~\cite{ATLAS:2019lng}\\&&\texttt{ATLAS-1712-08119($36.1fb^{-1}$)}~\cite{ATLAS:2017vat}\\&&\texttt{CMS-SUS-16-048($35.9fb^{-1}$)}~\cite{CMS:2018kag}\\\\
			\multirow{3}{*}{$\tilde{\chi}_{2}^{0}\tilde{\chi}_{1}^{\pm}+\tilde{\chi}_{1}^{\pm}\tilde{\chi}_{1}^{\mp}+\tilde{\chi}_{1}^{\pm}\tilde{\chi}_{1}^{0}$}&\multirow{3}{*}{$2\ell + nj(n\geq0) + \text{E}_\text{T}^{\text{miss}}$}&\texttt{ATLAS-1911-12606($139fb^{-1}$)}~\cite{ATLAS:2019lng}\\&&\texttt{ATLAS-1712-08119($36.1fb^{-1}$)}~\cite{ATLAS:2017vat}\\&&\texttt{CMS-SUS-16-048($35.9fb^{-1}$)}~\cite{CMS:2018kag}\\\hline 
\end{tabular}}
\end{table}

\bibliographystyle{CitationStyle}
\bibliography{LianRef}

\end{document}